\documentclass[prd,twocolumn,showpacs,showkeys,preprintnumbers,floatfix,
  nofootinbib,superscriptaddress]{revtex4-1}
\usepackage{amsfonts} 
\usepackage{amssymb} 
\usepackage{amsmath} 
\usepackage{subfigure} 
\usepackage{graphicx} 
\usepackage{array} 
\usepackage{dcolumn} 
\usepackage{bm} 
\usepackage{latexsym} 
\usepackage{longtable} 
\usepackage{hyperref} 
\usepackage{multirow}
\usepackage{color}
\usepackage{bbold}
\usepackage{lineno} 
\usepackage{verbatim}

\usepackage[section]{placeins}
\usepackage{morefloats}

\graphicspath{{./figs/}}

\newcommand{\SUSY}{\text{SUSY}}
\newcommand{\Cont}{\text{Cont}}
\newcommand{\Lat}{\text{Lat}}

\newcommand{\PQ}{\text{PQ}}
\newcommand{\PQA}{\text{PQA}}
\newcommand{\PQB}{\text{PQB}}
\newcommand{\SPQ}{\text{SPQ}}

\newcommand{\MeV}{\mathop{\rm MeV}\nolimits}
\newcommand{\GeV}{\mathop{\rm GeV}\nolimits}
\newcommand{\fm}{\mathop{\rm fm}\nolimits}

\newcommand{\overbar}[1]{\mkern 1.5mu\overline{\mkern-1.5mu#1\mkern-1.5mu}\mkern 1.5mu}

\begin{document}

\title{Toolkit for staggered $\Delta S=2$ matrix elements}
\author{Jongjeong Kim}
%
%
\affiliation{
  Lattice Gauge Theory Research Center, FPRD, and CTP, \\
  Department of Physics and Astronomy,
  Seoul National University, Seoul, 151-747, South Korea
}
\author{Weonjong Lee}
\email[E-mail: ]{wlee@snu.ac.kr}
%
%
%
\affiliation{
  Lattice Gauge Theory Research Center, FPRD, and CTP, \\
  Department of Physics and Astronomy,
  Seoul National University, Seoul, 151-747, South Korea
}
\author{Jaehoon Leem}
%
%
\affiliation{
  Lattice Gauge Theory Research Center, FPRD, and CTP, \\
  Department of Physics and Astronomy,
  Seoul National University, Seoul, 151-747, South Korea
}
\author{Stephen R. Sharpe}
\email[E-mail: ]{srsharpe@uw.edu}
%
%
\affiliation{
  Physics Department,
  University of Washington,
  Seattle, WA 98195-1560, USA
}
\author{Boram Yoon}
%
%
\affiliation{
  Los Alamos National Laboratory, MS B283, P.O. Box 1663,
  Los Alamos, NM 87545, USA
}
\collaboration{SWME Collaboration}
\date{\today}
\begin{abstract}
A recent numerical lattice calculation of
the kaon mixing matrix elements of general $\Delta S=2$ four-fermion
operators using staggered fermions relied on two auxiliary theoretical
calculations. Here we describe the methodology and present the results of
these two calculations.
The first concerns one-loop matching coefficients between
staggered lattice operators and the corresponding continuum operators.
Previous calculations with staggered fermions have used a non-standard
regularization scheme for the continuum operators, and here we
provide the additional matching factors needed to connect to the
standard regularization scheme. This is the scheme in which two-loop
anomalous dimensions are known.
We also observe that all previous calculations of this operator matching
using staggered fermions have overlooked one matching step in the continuum.
This extra step turns out to have no impact on three of the five operators
(including that relevant for $B_K$), but does affect the other two operators.
The second auxiliary calculation concerns
the two-loop renormalization group (RG)
evolution equations for the $B$-parameters of the $\Delta S=2$ operators.
For one pair of operators, the
standard analytic solution to the two-loop RG equations
fails due to a spurious singularity introduced by
the approximations made in the calculation.
We give a non-singular expression derived using analytic continuation,
and check the result using a numerical solution to the RG equations.
We also describe the RG evolution for
``golden'' combinations of $B$-parameters,
and give numerical results for RG evolution matrices
needed in the companion lattice calculation.
\end{abstract}
\maketitle
%
%
%


\allowdisplaybreaks


\section{Overview \label{sec:intr}}
There have been several recent lattice calculations 
of the kaon mixing matrix elements of 
all $\Delta S=2$ operators appearing in a
general theory of physics beyond the standard model (BSM)\cite{%
Boyle:2012qb,Bertone:2012cu,Bae:2013tca,Carrasco:2013jaa,Bae:2013mja,Lytle:2013oqa}.
These matrix elements are needed in order to use the experimental
results for $\varepsilon_K$ and $\Delta M_K$
to constrain the parameters of models of new physics. 
As members of the SWME collaboration, we have been involved in a
calculation using improved staggered fermions, which recently
presented results in Refs.~\cite{Bae:2013tca,Bae:2013mja}.
These results relied on two auxiliary theoretical calculations, and the
purpose of this paper is to present the details and results of these
calculations.

The first auxiliary calculation concerns the matching between the
continuum operators whose matrix elements we desire and the lattice
operators whose matrix elements we calculate.
We use one-loop perturbative matching.
The requisite one-loop calculations have been done in Ref.~\cite{Kim:2011pz},
but only
using a non-standard continuum scheme for defining four-fermion operators.
This scheme, introduced in Ref.~\cite{Gupta:1996yt}, 
has attractive properties under Fierz transformations, 
but has not been adopted in the continuum literature.
Instead, the standard continuum scheme is that used in Ref.~\cite{Buras:2000if}
to calculate the two-loop anomalous dimensions for the complete set
of $\Delta S=2$ operators.
This scheme differs from that of Ref.~\cite{Gupta:1996yt} in the choice of
evanescent operators.
Since we need the two-loop anomalous dimensions in order to evolve lattice
results to a common scale, it is necessary
to match the lattice operators to the standard continuum scheme.
Thus we have augmented the results of Ref.~\cite{Kim:2011pz} by calculating the
matching factor between the two continuum schemes.

Undertaking this relatively straightforward task, we have uncovered a
conceptual error in previous staggered perturbative matching calculations
for four-fermion operators~\cite{Sharpe:1993ur,Lee:2003sk,Kim:2011pz}.
It turns out that the matching factors obtained in these works connect
the lattice operators to continuum operators which
are non-standard not only because of the choice of scheme just described,
but also because of an additional finite correction.
Technically, this arises because
an additional continuum-to-continuum matching step is required.
In general this leads to a correction beginning at one-loop order. 
Since this point is of more general interest for applications 
using staggered fermions, we explain it in some detail.

It turns out that the
additional matching corrections vanish
for three of the five operators which arise in a general BSM theory.
In particular, previous results for the standard-model $B_K$ operator
are unaffected.
We also stress that the results for all five
operators presented in Refs.~\cite{Bae:2013tca,Bae:2013mja}
do include the correct matching factors.

The second auxiliary calculation concerns the renormalization group (RG)
running in the continuum. Our lattice calculation needs RG evolution to
convert results obtained at the lattice scale $1/a$ to a standard
scale such as $2\GeV$. 
Although this might appear to be a standard calculation, there are two
complications which arise.
First, the standard expressions for two-loop running
break down for one pair of operators,
due to a spurious singularity. In this case one can either use the
analytic continuation method of Ref.~\cite{Adams:2007tk}, 
or simply solve the RG equations numerically. 
We have compared these approaches, and present numerical
results for the evolution matrices.
The second complication is
that, in our lattice calculation, we make use of particular ``golden'' ratios
or products of $B$-parameters which are chosen to have simpler chiral
extrapolations~\cite{Becirevic:2004qd,Bailey:2012wb}. 
Here we present the formulae for RG evolution of these combinations.

A further reason for presenting our RG running factors is that there is some
disagreement between the results for BSM matrix elements of our work and
those of 
Refs.~\cite{Boyle:2012qb,Bertone:2012cu,Carrasco:2013jaa,Lytle:2013oqa}. 
Thus it is useful to present the technical details of our work so
as to facilitate a more thorough comparison.

This paper is organized as follows.
In Sec.~\ref{sec:op}, we recall the relevant
$\Delta S=2$ operators and define the corresponding $B$-parameters.
The method for calculating the one-loop matching factors 
is described in Section~\ref{sec:matching}, and
final results are presented.
The issues arising in RG evolution are described in
Section~\ref{sec:rg_evol}.
%
%
We include three appendices.
Appendix~\ref{app:sec:match} provides the technical details of the
calculation of the matching factors,
Appendix~\ref{app:sec:anomdim} collects results for anomalous dimensions,
and Appendix~\ref{app:sec:num} gives numerical results for evolution kernels.

\section{Continuum $\Delta S=2$ operators and $B$-parameters}
\label{sec:op}

The $\Delta S = 2$ effective Hamiltonian has the general form
\begin{equation}
  {\cal H}_\text{eff}^{\Delta S=2}
  = \sum_i C_i(\mu) {Q}_i(\mu) \,,
  \label{eq:Heff}
\end{equation}
where the $Q_i$ form a basis of $\Delta S=2$ four-fermion operators, 
and the $C_i$ are Wilson coefficients.
This form holds both in the standard model (SM) and in a general BSM theory,
and arises after heavy particles are integrated out.
Contributions from operators of higher dimension are neglected.
Both $C_i$ and $Q_i$ depend on the renormalization scale $\mu$,
as is displayed explicitly.
They also have an implicit dependence on the regularization scheme
used to define the operators. 
This could either be a continuum scheme or some form of lattice regularization.
The scheme and scale dependence cancels in ${\cal H}_\text{eff}$,
and using this one can determine how the $C_i$ depend on the scheme
and on $\mu$. Determining the relationship between the $C_i$ in different
schemes and at different scales is the focus of this paper.

We first consider the form of the operators that appear in continuum
regularization. These will be given a superscript ``Cont''.
In the SM, the left-handed couplings of the $W$-boson imply that only
a single operator has a non-vanishing Wilson coefficient, namely that
with ``left-left'' spin structure:
\begin{align}
{Q}^\text{Cont}_{1}\equiv {Q}^\text{Cont}_{K} &=
 [\bar{s}^a \gamma_\mu L d^a] 
 [\bar{s}^b \gamma_\mu L d^b] \,.
 \label{eq:Q1}
\end{align}
Here $L=(1\!-\!\gamma_5)$,
$a$ and $b$ are color indices, and repeated indices are summed. 
We work in Euclidean space throughout.
In a general BSM theory, four other operators appear
in addition to Eq.~\eqref{eq:Q1}. These can be chosen to be
\begin{align}
 {Q}^\text{Cont}_{2} &= 
 [\bar{s}^a L d^a] [\bar{s}^b L d^b] \,,  
 \label{eq:Q2}\\
 {Q}^\text{Cont}_{3} &=
 [\bar{s}^a \sigma_{\mu\nu} L d^a]
 [\bar{s}^b \sigma_{\mu\nu} L d^b] \,,
 \label{eq:Q3}\\
 {Q}^\text{Cont}_{4} &= 
 [\bar{s}^a L d^a] [\bar{s}^b R d^b] \,,
 \label{eq:Q4}\\
 {Q}^\text{Cont}_{5} &= 
 [\bar{s}^a \gamma_\mu L d^a] 
  [\bar{s}^b \gamma_\mu R d^b] \,,
  \label{eq:Q5}
\end{align}
where $R=(1\!+\!\gamma_5)$ and $\sigma_{\mu\nu} = [\gamma_\mu, \gamma_\nu]/2$.
This is essentially the basis given in Ref.~\cite{Buras:2000if},
which we call the ``Dirac basis''.\footnote{%
Specifically, our operators are related to those of Ref.~\cite{Buras:2000if}
by $Q^\text{Cont}_{2,3} = 4 Q_{1,2}^{SLL}$ 
and $Q^\text{Cont}_{4,5} = 4 Q_{2,1}^{LR}$. 
The factor of four arises because we use
$(1\pm\gamma_5)$ instead of $(1\pm\gamma_5)/2$
in order to simplify some subsequent results.
This factor cancels in suitably defined $B$-parameters and 
in anomalous dimensions.
We have also reordered the ``$LR$'' operators.}
A complete definition also requires a choice of basis for the
evanescent operators, i.e. those which appear when
one extends from $4$ to $D=4-2\epsilon$ dimensions.
We use the choice of Ref.~\cite{Buras:2000if}. 
This is the scheme in which the two-loop anomalous dimensions have
been calculated.

The list of operators given above is, in fact, incomplete.
Three more operators can appear---those obtained from
$Q_{1,2,3}^\text{Cont}$ by interchanging $L$ and $R$.
We do not consider these operators separately because we are
ultimately interested in the positive parity parts of all operators, which
are the same for both left- and right-handed operators.
Only the positive parity parts contribute to the $K^0-\overbar{K}^0$ mixing
matrix elements.
Furthermore, the matching of the right-handed operators
to the corresponding lattice operators involves identical
coefficients as for the left-handed operators, and the RG running is
also identical.

It is useful in lattice calculations to determine dimensionless $B$-parameters
rather than matrix elements. For the Dirac basis operators, these are
\begin{align}
  B_1 &= B_K =
  \frac{\langle \overbar{K}^0 \vert Q^\text{Cont}_i \vert K^0 \rangle}
       {N_1 \langle \overbar{K}^0 \vert \overline{s} 
       \gamma_\mu \gamma_5 d\vert 0 \rangle
        \langle 0 \vert \bar{s}\gamma_\mu \gamma_5 d \vert K^0 \rangle}  
\label{eq:def-B_K}\\
  B_i &= 
  \frac{\langle \overbar{K}^0 \vert Q^\text{Cont}_i \vert K^0 \rangle}
       {N_i \langle \overbar{K}^0 \vert \overline{s}\gamma_5 d\vert 0 \rangle
        \langle 0 \vert \bar{s} \gamma_5 d \vert K^0 \rangle} 
       \qquad \text{for } i=2,3,4,5
\label{eq:def-B_i}
\end{align}
where $N_j = \left\{\dfrac{8}{3}, \dfrac{5}{3}, 4, -2, \dfrac{4}{3}\right\}$
for $j=1,2,3,4,5$, respectively.
The denominators are obtained using the vacuum insertion approximation
though only keeping the leading terms in the SU(3) chiral limit.
We stress that these $B$-parameters are simply useful intermediate quantities,
with their precise definition being immaterial as long as one uses the same
definition throughout.

An alternative to the Dirac basis is the ``SUSY basis'' 
of Ref.~\cite{Gabbiani:1996hi}:
\begin{align}
  \mathcal{O}^\text{Cont}_{1} &= Q_1^\text{Cont}\,, \\
 \mathcal{O}^\text{Cont}_{2} &= Q_2^\text{Cont}\,,\\
 \mathcal{O}^\text{Cont}_{3} &=
 [\bar{s}^a L d^b] [\bar{s}^b L d^a] \,,
\\
 \mathcal{O}^\text{Cont}_{4} &= Q_4^\text{Cont}\,,\\
 \mathcal{O}^\text{Cont}_{5} &=
 [\bar{s}^a L d^b] [\bar{s}^b R d^a] \,.
\end{align}
This has been used, for example, in the lattice calculations of 
Refs.~\cite{Boyle:2012qb,Bertone:2012cu,Carrasco:2013jaa,Lytle:2013oqa}.
The corresponding $B$-parameters are defined as in
Eqs.~(\ref{eq:def-B_K}) and (\ref{eq:def-B_i}), except with
$N_3=-1/3$ and $N_5=-2/3$.
In four dimensions one can relate the two bases using Fierz
transformations, while in $D \neq 4$ dimensions the relation involves 
additional evanescent operators:
\begin{align}
Q_3^\text{Cont} &= 
4 {\cal O}^\text{Cont}_2 + 8 {\cal O}^\text{Cont}_3 + \text{evanescent} \,,
\label{eq:Q3oldvsnew}\\
Q_5^\text{cont} &= - 2 {\cal O}^\text{Cont}_5 + \textrm{evanescent}\,.
\label{eq:Q5oldvsnew}
\end{align}
A key point, however, is that the  way the SUSY basis operators
are defined in 
Refs.~\cite{Boyle:2012qb,Bertone:2012cu,Carrasco:2013jaa,Lytle:2013oqa}
is by using the {\em four-dimensional} Fierz transform to relate them
to the Dirac basis. It is in the latter basis that the
evanescent operators are defined and in which RG running is done.
This means that the $B$-parameters in the two bases can be related
simply using the $D=4$ results.
In particular, $B_i^\text{SUSY}=B_i$ for $i=1$, $2$, $4$, and $5$,
while
\begin{equation}
B_3^\SUSY= -\frac32 B_3 + \frac52 B_2\,.
\end{equation}
The latter result follows from the $D=4$ relation
\begin{equation}
{\cal O}^\SUSY_3 = \frac{Q_3-4Q_2}{8}\,,
\end{equation}
obtained by inverting Eq.~\eqref{eq:Q3oldvsnew} in $D=4$.

\section{One-loop Matching}
\label{sec:matching}

As noted in the Introduction, one-loop matching
calculations with staggered fermions~\cite{Sharpe:1993ur,Lee:2003sk,Kim:2011pz}
use different continuum operators than those discussed
in the previous section.
The difference is twofold: 
the use of a different basis of evanescent operators and
a missing matching step.
In this section we describe how to change the previous
calculations in order to match to the desired continuum
operators. The key is to understand the impact of
the extra tastes that come with staggered fermions.

It turns out that the just-mentioned
differences in continuum operators have no impact
on the one-loop matching factors for the continuum
operators $Q_1^\text{Cont}$, $Q_4^\text{Cont}$ and $Q_5^\text{Cont}$.
Thus the matching factors for these operators obtained in 
Ref.~\cite{Kim:2011pz} are correct.
Why this is the case will become clear only when the
analysis is complete.
Given this result, we couch our
discussion in terms of the operators $Q_2^\text{Cont}$ and $Q_3^\text{Cont}$, 
for which the differences do lead to changes in the matching factors.

\subsection{Staggered Complications}
\label{subsec:complications}

In a lattice calculation with staggered fermions, one must deal
with the fact that each lattice field yields four degenerate tastes
in the continuum limit. For sea quarks this is 
done by taking the fourth root of the fermion determinant. 
This prescription is not controversial in perturbation theory, where
it is implemented by dividing each quark loop by a factor of four.
In fact, for the matching factors we consider, quark loops do not enter
until two-loop order so we will not need this prescription for
our one-loop calculation.

For the valence quarks, on the other hand, one must account for the fact
that the lattice theory has more degrees of freedom than QCD. This means
that, even in the continuum limit (where taste symmetry is restored) the
lattice theory is {\em different} from QCD. In particular, it is necessarily
a partially quenched (PQ) theory. Although ``rooting'' ensures that the
$\beta$-function agrees with that of QCD, the matching of operators, where
rooting is not an option, is more complicated.

To understand this in more detail, consider the matrix element of
$Q_{2}^\Cont$ [Eq.~(\ref{eq:Q2})] between an external kaon and antikaon
in QCD. Both particles are destroyed/created by a local,
color-singlet operator of the form
$\bar d^a\gamma_5 s^a$. The matrix element involves two types of Wick
contractions, one in which the fields in the external operator are both
contracted with the $\bar s$ and $d$ in a single bilinear, and the other in
which the external fields are contracted with an $\bar s$ from one bilinear
and a $d$ from the other. 
In the first type of contraction the color indices form two loops,
while in the second they form a single loop.
Thus we refer to them respectively as ``two color-loop'' 
and ``one color-loop'' contractions.\footnote{%
This classification into two types of contraction holds also in perturbation
theory (PT),
although the description in terms of color-loops is less appropriate.
This is because, in PT, one uses external quark fields with uncontracted
Dirac and color indices and having definite momentum rather
than pseudoscalar, color-singlet kaon operators. Specifically, one uses
$\bar d^a_\alpha(p_1) s^b_\beta(p_2) \bar d^c_\gamma(p_3) s^d_\gamma(p_4)$
in QCD.
One can, however, still group the fields into two $\bar d s$
pairs in an unambiguous (although arbitrary)
way using the external indices and/or momenta as labels,
and then define one and two color-loop contractions relative
to those pairings.}
At tree-level, where one can work in four
dimensions, the one color-loop contraction can be rewritten 
by doing a Fierz transformation on the operator:
\begin{align}
Q_{2}^\Cont &\stackrel{D=4}{=} 
-\frac12 [\bar{s}^a L d^b] [\bar{s}^b L d^a] 
\nonumber\\
&\quad + \frac18
 [\bar{s}^a \sigma_{\mu\nu}L d^b] 
 [\bar{s}^b \sigma_{\mu\nu}L d^a] \,.
\label{eq:Q1SLL_Fierz}
\end{align}
In this form, the one color-loop contraction now has the fields
in each external operator contracted with those in a single bilinear.
Note that the Fierz-transformed forms involve the same Dirac structures 
as in $Q_{2,3}^\Cont$, but with color indices contracted differently. 

We next consider the analogous operators in the continuum
limit of the staggered theory. In this theory we have
fields $S$ and $D$, where upper case is used to indicate
that there are four tastes of each of the valence quarks,
so that $S$ and $D$ are vectors with an implicit taste index.
A possible choice of operator to match with $Q_2^\Cont$ is then
\begin{equation}
[\bar{S}^a (L \otimes \xi_5) D^a] 
[\bar{S}^b (L \otimes \xi_5) D^b] 
\,.
\label{eq:naivestagop}
\end{equation}
Here the second matrix in each tensor product indicates the taste
matrix.  We have chosen the bilinears to have ``Goldstone'' taste,
since that is what is done in actual lattice calculations, but we
stress that the problem we are about to explain occurs for any choice
of taste. If we now take the matrix element of this operator between a
kaon destroyed by $\bar{D} \gamma_5\otimes\xi_5 S$ and an antikaon
created by an operator of the same form, there will again be two types
of Wick contraction.  At tree-level, the two color-loop contraction
will be the same as that for $Q_2^\Cont$ in QCD, aside from an overall
taste factor of $N_T^2$, where $N_T=4$.  (This arises because there
are two ``taste-loops'', in each of which all four tastes can flow.)
To evaluate the one color-loop contraction at tree-level we can
Fierz-transform the operator so that the contraction involves two
taste loops.  This now requires simultaneous Fierz transformations on
Dirac and taste indices. The former transform as in
Eq.~(\ref{eq:Q1SLL_Fierz}), while the taste transformation is
\begin{equation}
\xi_5 \cdot \xi_5 \longrightarrow
\sum_F \frac{{\rm tr}(\xi_5 \xi_F \xi_5 \xi_F)}{N_T^2}
\;\xi_F \cdot \xi_F
\,,
\end{equation}
with $F$ being summed over all sixteen tastes.
Upon contraction with the external kaons of taste $\xi_5$,
only the $F=5$ term contributes.
This comes with a ``Fierz factor'' of ${\rm tr}{\bf 1}/N_T^2=1/N_T$
as well as the overall factor of $N_T^2$.
Thus the one color-loop contraction at tree-level
is the same that for $Q_2^\Cont$ in QCD, aside from a taste factor 
of $N_T$.
We now can see the key problem: the two types of contraction come with 
{\em different} taste factors compared to the QCD operator.
Thus, even with an overall rescaling, the entire matrix elements cannot match.
This is the inevitable consequence of the presence of the additional tastes.

This problem has been recognized since the
first calculation of matrix elements using staggered 
fermions~\cite{Sharpe:1986xu}, and the solution adopted has been to
match Wick contractions rather than operators.
This solution is explained in Ref.~\cite{Kilcup:1997ye},
but, as noted above, is incomplete.
In the next few subsections we give the complete description,
which involves a sequence of four matching steps.

\subsection{First Matching Step: QCD to PQQCD}
\label{subsec:step1}

In the first step we match from QCD to a partially quenched extension of QCD
in which there are two degenerate valence strange quarks, $s_1$ and $s_2$,
and two degenerate valence down quarks, $d_1$ and $d_2$. 
The sea-quark composition is the same as in QCD, 
and we consider this theory only in the continuum.
In this paper we refer to this specific theory as PQQCD.
At this stage there is no taste degree of freedom, so this
is {\em not} the continuum limit of a staggered lattice theory.
We regulate this theory using dimensional regularization,
using an NDR scheme in which evanescent operators are generalized from
QCD to the PQ theory in the simplest way (as discussed below).
The reason for introducing this theory
is that it allows us to separate the
two types of Wick contraction without needing to deal with the complications
arising from the additional tastes.

Consider the matrix element in PQQCD of 
\begin{equation}
 Q_{2,II}^\PQ = 2
 [\bar{s}_1^a L d_1^a] 
 [\bar{s}_2^b L d_2^b] 
\label{eq:Q2IIPQ}
\end{equation}
between a $K_1^{0}$ created by $\bar d_1\gamma_5 s_1$
and a $\overbar{K}^{0}_2$ destroyed by $\bar d_2\gamma_5 s_2$.
This matrix element is identical, diagram by diagram in PT, to the two color-loop
Wick contractions of $Q_2^\Cont$ between an external kaon and antikaon in QCD.
The factor of 2 in Eq.~(\ref{eq:Q2IIPQ}) is needed because, in QCD,
each external operator can be contracted with either bilinear, while in PQQCD
this is not possible.
Because the matching is with the two color-loop contraction in QCD,
we label $Q_{2,II}^\PQ$ with the additional subscript $II$.

This diagram by diagram equality in fact holds much more generally.
If one uses the external fields
$\bar d^a_{\alpha}(p_1) s^b_{\beta}(p_2) 
 \bar d^c_{\gamma}(p_3) s^d_{\delta}(p_4)$ in QCD 
(with $\alpha-\delta$ Dirac indices), and
keeps only the contractions in which the fields with momenta $p_1$ and $p_2$
are connected to the same bilinear
(so that the fields with momenta $p_3$ and $p_4$ are connected to the other bilinear)
then the matrix element agrees exactly with that in PQQCD
with external fields
$\bar d^a_{1,\alpha}(p_1) s^b_{1,\beta}(p_2) 
 \bar d^c_{2,\gamma}(p_3) s^d_{2,\gamma}(p_4)$.
This holds for all values of the external Dirac and color indices, 
and for all choices of the momenta $p_i$.

In a similar way, the one color-loop contractions of $Q_2^\Cont$ in QCD
matches exactly to the PQQCD matrix element of
\begin{equation}
 Q_{2,IA}^\PQ = 2
 [\bar{s}_1^a L d_2^a] 
 [\bar{s}_2^b L d_1^b] \,.
\label{eq:Q2IAPQ}
\end{equation}
Here, the subscript ``$I$'' indicates matching with a one color-loop
contraction, while ``$A$'' distinguishes the operator from a
similar one introduced below.
Note that this operator differs from $Q_{2,II}^\PQ$ only
by the interchange $d_1\leftrightarrow d_2$ between the bilinears,
while keeping each bilinear a color singlet.
In particular, no Fierz transformation has been done on $Q_2^\Cont$,
so that the exact matching holds for $D=4-2\epsilon$.

Repeating this exercise for $Q_3^\Cont$ one finds that the
PQ operator corresponding to its two and one color-loop contractions are,
respectively,
\begin{align}
 Q_{3,II}^\PQ &= 2
 [\bar{s}_1^a \sigma_{\mu\nu} L d_1^a] 
 [\bar{s}_2^b \sigma_{\mu\nu} L d_2^b]\,,
\label{eq:Q3IIPQ}
\\
 Q_{3,IA}^\PQ &= 2
 [\bar{s}_1^a \sigma_{\mu\nu} L d_2^a] 
 [\bar{s}_2^b \sigma_{\mu\nu} L d_1^b]\,.
\label{eq:Q3IAPQ}
\end{align}

To write a matching equation involving operators we form the
linear combinations
\begin{equation}
Q_{j,\pm}^\PQ = Q_{j,II}^\PQ \pm Q_{j,IA}^\PQ \qquad (j=2,3)
\,.
\label{eq:QPQpm}
\end{equation}
Our claim is that, for matrix elements involving the external
operators described above, we have, to all orders in PT
\begin{equation}
Q_j^\Cont \cong Q_{j,+}^\PQ \qquad (j=2,3)\,.
\label{eq:matchContPQ}
\end{equation}
This is our first matching equation.
The two operators on the r.h.s. are needed to obtain both Wick contractions
of the operator on the l.h.s.
The symbol ``$\cong$'' indicates that this is not a true operator matching,
but rather that the matrix elements {\em of the type described above}
agree between the two theories.
This is sufficient for our purposes since these are the matrix elements
of interest.

The difference operators $Q_{j,-}^\PQ$ in Eq.~(\ref{eq:QPQpm}) do not
play a role in the matching to $Q_j^\Cont$.
In fact, they are PQQCD operators with
no counterparts in the $\Delta S=2$ sector of QCD.
We will use them, however, in the next stage of the calculation.

As already noted, when doing a perturbative calculation of
the matrix elements described above, one
encounters additional, evanescent operators which must be
dealt with in order to renormalize the matrix elements.
These are local operators with Dirac structures that vanish
when $D=4$. In order for the above-described exact matching
to hold after renormalization, evanescent operators must be
treated in the same way in both QCD and PQQCD.
Doing so is, in fact, completely straightforward, since
the treatment in QCD is already done contraction by contraction.
Concrete examples of this statement are given in the
explicit calculation of Appendix~\ref{app:sec:match}.

We stress that, although the exact equality of matrix elements
described in this subsection is almost trivial,
it is nevertheless useful in order to set-up the next, non-trivial,
stage of the matching.
We also note that our argument is a minor adaptation of that 
used in Ref.~\cite{Buras:2000if}
to show how the anomalous dimensions of
$\Delta F=1$ operators with flavor $\bar s d \bar u c$
can be related to those of $\Delta S=2$ operators.

\subsection{Second Step: Basis Change in PQQCD}
\label{subsec:step2}

At this stage we have succeeded in exactly converting the 
desired QCD calculation into one in PQQCD. 
The next step is to change the operator basis in PQQCD.
Essentially, we are doing a Fierz transform on the operators
which match with one color-loop contractions in QCD, but
taking into account the failure of Fierz transforms away from $D=4$.
This step is useful since the new basis in PQQCD
matches straightforwardly onto the lattice theory.

We collect the operators discussed in the previous subsection
into a vector,
\begin{equation}
\overrightarrow{{\cal O}^\text{PQA}}
= \{ Q_{2,+}^\PQ, \;Q_{3,+}^\PQ,\;Q_{2,-}^\PQ,\;Q_{3,-}^\PQ \}
\,.
\label{eq:basisPQA}
\end{equation}
We will change from this basis to
\begin{equation}
\overrightarrow{{\cal O}^\text{PQB}} = \{ Q_{2,I}^\PQ,
\;Q_{2,II}^\PQ,\;Q_{3,I}^\PQ,\;Q_{3,II}^\PQ \}
\,.
\label{eq:basisPQB}
\end{equation}
Here $Q_{2,II}^\PQ$ and $Q_{3,II}^\PQ$ are defined
in Eqs.~(\ref{eq:Q2IIPQ}) and (\ref{eq:Q3IIPQ}) above, while
\begin{align}
Q_{2,I}^\PQ \equiv {\cal O}_1^\PQB
&= 2 [\bar{s}_1^a L d_1^b] 
                       [\bar{s}_2^b L d_2^a] 
\label{eq:Q2IPQ}
\\
Q_{3,I}^\PQ \equiv {\cal O}_3^\PQB
&= 2 [\bar{s_1}^a \sigma_{\mu\nu} L d_1^b] 
                      [\bar{s_2}^b \sigma_{\mu\nu} L d_2^a] 
\,.
\label{eq:Q3IPQ}
\end{align}
These are the two operators one obtains from $Q_{2,IA}^\PQ$ 
and $Q_{3,IA}^\PQ$ by
interchanging $d_2^a$ and $d_1^b$.
For $D=4$ such an interchange is brought about by a Fierz transformation,
which also effects the Dirac structure.
Specifically, we have
\begin{eqnarray}
Q_{2,IA}^\PQ &\stackrel{D=4}{=}&
-\frac12 Q_{2,I}^\PQ+\frac18 Q_{3,I}^\PQ 
\label{eq:FierzPQ2}
\\
Q_{3,IA}^\PQ &\stackrel{D=4}{=}&
6 Q_{2,I}^\PQ + \frac12 Q_{3,I}^\PQ
\,.
\label{eq:FierzPQ3}
\end{eqnarray}
so that
\begin{equation}
{\cal O}_k^\text{PQA}
\stackrel{D=4}{=}
R_{k\ell} 
{\cal O}_\ell^\text{PQB}
\,,
\label{eq:lintrans}
\end{equation}
with 
\begin{equation}
R = \left(\begin{array}{cccc} -\frac12 & 1 & \frac18 & 0\\
                              6 & 0 & \frac12 & 1 \\
                              \frac12 & 1 & -\frac18 & 0\\
                              -6 & 0 & -\frac12 & 1
    \end{array}\right)\,.
\end{equation}
This means that tree-level matrix elements with 
the external fields described in the previous subsection 
will be related by the same linear transformation
\begin{equation}
\langle {\cal O}_k^\text{PQA}\rangle^\text{(0)} =
R_{k\ell} 
\langle {\cal O}_\ell^\text{PQB}\rangle^\text{(0)}
\,,
\label{eq:lintransME0}
\end{equation}
Here the superscript indicates the order in $\alpha$.

This simple relation does not hold beyond tree level,
since the Fierz transforms (\ref{eq:FierzPQ3}) fail for $D=4-2\epsilon$.
This is a standard situation in renormalization theory,
explained clearly, for example, in Ref.~\cite{Buras:2000if}.
The basis must be extended to include evanescent operators.
At one-loop order, all one needs, in fact, is a set of projectors
which pick out the components of the desired operators from
expressions in $4-2\epsilon$ dimensions. 
The projectors for the operators we consider
have been given in Ref.~\cite{Buras:2000if},
and are conveniently summarized in Ref.~\cite{Buras:2012fs}.
The general result is that
one-loop anomalous dimensions are the same in the two bases
(once the linear transformation given in Eq.~(\ref{eq:lintrans})
is taken into account),
while finite parts of one-loop matrix elements 
(and correspondingly two-loop anomalous dimensions) can be different.
This is thus an example of non-trivial operator matching, although
simplified because both sets of operators are defined
in the same theory. 

To determine the one-loop matching between the bases one
calculates the one-loop matrix elements with the same external
fields and equates them.
After renormalization, these matrix elements take the forms
\begin{align}
\langle {\cal O}_k^\text{PQA} \rangle^\text{(1)}
&= Z^\text{PQA}_{k\ell} 
\langle {\cal O}_\ell^\text{PQA} \rangle^\text{(0)}
\,, \\
Z^\text{PQA}_{k\ell} &= \delta_{k\ell}
+ \frac{\alpha}{4\pi}\left[
\gamma^\PQ_{k\ell} \log(\lambda/\mu) + C_{k\ell}^\text{PQA}\right]
\,,
\label{eq:PQAmatch}
\end{align}
and 
\begin{align}
\langle {\cal O}_k^\text{PQB} \rangle^\text{(1)}
&= Z^\text{PQB}_{k\ell} 
\langle {\cal O}_\ell^\text{PQB} \rangle^\text{(0)} \,, \\
 Z^\text{PQB}_{k\ell}  &=
 \delta_{k\ell} + \frac{\alpha}{4\pi}\left[
(R^{-1}\gamma^\PQ R)_{k\ell} \log(\lambda/\mu) + C_{k\ell}^\text{PQB}\right]
\,.
\label{eq:PQBmatch}
\end{align}
Here $\lambda$ is an infrared cut-off (for which we use a gluon mass),
$\mu$ the renormalization scale,
and $\gamma^\PQ$ the one-loop anomalous dimension in the PQA basis.
The factors of $R^{-1}$ and $R$ in the result for
$Z^\text{PQB}$ are needed so that the one-loop anomalous
dimensions match once the linear transformation
(\ref{eq:lintrans}) is taken into account.
$C^\text{PQA}$ and $C^\text{PQB}$ are the finite parts of the
one-loop result.

Equating $\langle {\cal O}_k^\text{PQA} \rangle^\text{(1)}$
with $R_{k\ell}\langle {\cal O}_\ell^\text{PQ} \rangle^\text{(1)}$
(a step that can be done in $D=4$ since the matrix elements have been
renormalized), and using (\ref{eq:lintransME0}), one finds
our second matching equation
\begin{equation}
{\cal O}_k^\text{PQA} \cong
\left\{ R_{k\ell} + \frac{\alpha}{4\pi} 
\left[(C^\text{PQA}R)_{k\ell} - (R C^{PQB})_{k\ell} \right]\right\} 
{\cal O}_\ell^\text{PQB}
\,.
\label{eq:matchPQ}
\end{equation}
The precise meaning of this equation is that the one-loop matrix elements
of the operators on the two sides agree, as long as one uses the
same definitions of evanescent operators as were used to determine the
matrices $C^{PQA}$ and $C^{PQB}$. We still use the symbol $\cong$,
although here the theory is the same on both sides of the matching,
because we want to allow for the possibility of using a different
renormalization scheme for the two bases.

We calculate the difference matrix $C^{PQA}R-RC^{PQB}$ in 
Appendix~\ref{app:sec:match}.
It is convenient to use different definitions
of evanescent operators for the PQA and PQB bases.
For the former we use the definitions of Ref.~\cite{Buras:2000if},
so that we are ultimately matching to an operator basis in which
we know the two-loop anomalous dimensions.
For the PQB basis, however, we adopt the NDR$'$ scheme,
which was introduced in Ref.\cite{Sharpe:1993ur}.
This is a convenient choice as it allows us to piggyback on 
previous one-loop calculations.
We stress, however, that even if we use the
definitions of Ref.~\cite{Buras:2000if} in both bases, the 
one-loop matching would be nontrivial.

\subsection{Third Step: PQQCD to Continuum Staggered Theory}
\label{subsec:step3}

The next step is to match to the theory
obtained in the continuum limit of a staggered lattice theory,
which we refer to as the ``SPQ'' theory (with S for staggered).
This differs from PQQCD by the presence of additional tastes. 
Specifically, this new theory has valence quarks $S_j$ and $D_j$,
with $j=1,2$, in addition to the (rooted) sea quarks.
As above, upper-case letters indicate the presence of four tastes.\footnote{%
The SPQ theory differs from the staggered theory discussed 
in Sec.~\ref{subsec:complications} in which there was only 
one $S$ and one $D$ quark.}

The operators we consider in the SPQ theory are simple
generalizations of those in the PQB basis in PQQCD.
They are obtained by replacing lower-case fields with their upper-case versions,
and inserting the taste matrix $\xi_5$.
For example,
\begin{eqnarray}
\lefteqn{Q_{2,I}^\PQ = 2 [\bar{s}_1^a L d_1^b] 
                       [\bar{s}_2^b L d_2^a] }
\nonumber \\
&\longrightarrow& 
2 [\bar{S}_1^a (L\otimes \xi_5) D_1^b] 
                       [\bar{S}_2^b (L \otimes \xi_5) D_2^a] 
\,.
\end{eqnarray}
In addition we will keep only the positive parity parts of the
operators, since these are the parts which contribute to
the $K^0-\overbar{K}^0$ matrix elements in which we are
ultimately interested. This has no impact on anomalous dimensions or
matching coefficients.
In this way we arrive at the basis
\begin{align}
{\cal O}^\SPQ_1 &= 
2 \Big([\bar{S}_1^a (\mathbf{1}\otimes \xi_5) D_1^b] 
  [\bar{S}_2^b (\mathbf{1}\otimes \xi_5) D_2^a] \nonumber \\
&\qquad +
 [\bar{S}_1^a (\gamma_5\otimes \xi_5) D_1^b] 
 [\bar{S}_2^b (\gamma_5\otimes \xi_5) D_2^a] \Big)
\label{eq:O1SPQ}
\\
{\cal O}^\SPQ_2 &= 
2\Big( [\bar{S}_1^a (\mathbf{1}\otimes \xi_5) D_1^a] 
  [\bar{S}_2^b (\mathbf{1}\otimes \xi_5) D_2^b] 
  \nonumber \\ &\qquad
+
  [\bar{S}_1^a (\gamma_5\otimes \xi_5) D_1^a] 
  [\bar{S}_2^b (\gamma_5\otimes \xi_5) D_2^b] \Big)
\label{eq:O2SPQ}
\\
{\cal O}^\SPQ_3 &= 
4 [\bar{S}_1^a (\sigma_{\mu\nu}\otimes \xi_5) D_1^b] 
  [\bar{S}_2^b (\sigma_{\mu\nu}\otimes \xi_5) D_2^a] 
\label{eq:O3SPQ}
\\
{\cal O}^\SPQ_4 &= 
4 [\bar{S}_1^a (\sigma_{\mu\nu}\otimes \xi_5) D_1^a] 
  [\bar{S}_2^b (\sigma_{\mu\nu}\otimes \xi_5) D_2^b] 
\label{eq:O4SPQ}
\,.
\end{align}
Note that for the ``tensor'' operators
${\cal O}^\SPQ_{3,4}$ there is only one term, since the
operators differing by a factor of $\gamma_5\cdot\gamma_5$ are identical, 
and can thus be combined.
This changes the overall factor from $2$ to $4$.

We now consider the matching between matrix elements
of the PQB basis operators in PQQCD
and those of the above-described operators in the SPQ theory.
In PQQCD we use the external operators $\bar d_1\gamma_5 s_1$
and $\bar d_2\gamma_5 s_2$, as already discussed in Sec.~\ref{subsec:step1}.
In the SPQ theory we use $\bar D_1(\gamma_5\otimes\xi_5) S_1$
and $\bar D_2(\gamma_5\otimes\xi_5) S_2$.
Thus only positive parity operators contribute to the matrix elements.
We now observe that, at any order in PT, the diagrams contributing
to the matrix elements of ${\cal O}^\PQB_k$ in the PQ theory
are identical to those contributing to the corresponding
matrix elements of ${\cal O}^\SPQ_k$ in the SPQ theory,
aside from the presence of the taste matrices $\xi_5\cdot \xi_5$.
Given the exact taste symmetry of the SPQ theory, however,
the extra taste factors lead only to an overall factor of $N_T^2$,
which can be removed by hand.
Thus, as long as we use the analogous choices for evanescent operators
in the two theories,\footnote{%
In practice, we use the NDR$'$ scheme for both theories.}
there is an exact matching of matrix elements.
We write this result as
\begin{equation}
{\cal O}^\PQB_k \cong {\cal O}^\SPQ_k\,,
\label{eq:matchPQSPQ}
\end{equation}
where we are stretching the meaning of ``$\cong$'' here to
include the provisos that taste factors are removed and only
a particular class of matrix elements is considered.
We also note that a consequence of this exact matching
is that anomalous dimensions agree to all orders.

\subsection{Final Step: Continuum to Lattice Staggered Theory} 
\label{subsec:final}

The final matching step is between the SPQ theory and the
lattice theory using improved staggered fermions.
This is a conventional matching between the same theory
regularized in two different ways: dimensional regularization,
with operators defined in the NDR$'$ scheme for the SPQ theory,
and lattice regularization, for which no issues of
evanescent operators arise.
For the operators we use in our numerical calculation, the
required one-loop matching has been done in Ref.~\cite{Kim:2011pz}.
The only subtlety is that, due to the breaking of taste symmetry
by lattice regularization, the basis of lattice operators
which mix with one another is much larger than that in the
continuum theory.
As in Ref.~\cite{Kim:2011pz}, we show here only the mixing with
operators having the same taste as those in the
SPQ theory, namely $\xi_5\cdot\xi_5$. These are the operators
used in present simulations. Dropping operators with other tastes
leads to an error suppressed by both $\alpha$ and by
a factor of $m_\pi^2/m_K^2$ or $m_\pi^2/(4\pi f_\pi)^2$~\cite{Bailey:2012wb}.

There are six lattice operators with taste
$\xi_5\cdot\xi_5$ that enter, and we collect these into a vector:
\begin{equation}
\overrightarrow{{\cal O}^\text{Lat}}=\{
{\cal O}^\text{Lat}_{S1},
{\cal O}^\text{Lat}_{S2},
{\cal O}^\text{Lat}_{P1},
{\cal O}^\text{Lat}_{P2},
{\cal O}^\text{Lat}_{T1},
{\cal O}^\text{Lat}_{T2}\}
\,.
\end{equation}
Here we are using the notation and definitions of Ref.~\cite{Kim:2011pz}.
The subscripts indicate, first, the
nature of the Dirac matrices (scalar, pseudoscalar or tensor)
and, second, the color contraction (one or two color-loops).
We do not repeat the details here.
The difference from the basis ${\cal O}^\SPQ$ of
Eqs.~(\ref{eq:O1SPQ}-\ref{eq:O4SPQ}) is that
the parts of ${\cal O}^{SPQ}_{1,2}$ with Dirac structures
$1\cdot 1$ and $\gamma_5\cdot\gamma_5$ have been separated
in the lattice operators.
This is required because they renormalize differently.

The tree-level relationship between the bases is
\begin{equation}
\langle {\cal O}_k^\text{SPQ}\rangle^\text{(0)} =
2 S_{km} 
\langle {\cal O}_m^\text{Lat}\rangle^\text{(0)}
\,,
\label{eq:lintransS}
\end{equation}
with $S$ the rectangular matrix 
\begin{equation}
S = \left(\begin{array}{cccccc} 
  1 & 0 & 1 & 0 & 0 & 0 \\
  0 & 1 & 0 & 1 & 0 & 0 \\
  0 & 0 & 0 & 0 & 4 & 0 \\
  0 & 0 & 0 & 0 & 0 & 4
      \end{array}\right)\,.
\end{equation}
The overall factor of $2$ in (\ref{eq:lintransS})
appears because the definition of
the lattice operators does not include the overall factors of $2$ that
appear in the continuum PQ operators
[see Eqs.~(\ref{eq:O1SPQ}) and (\ref{eq:O2SPQ})].
The factors of $4$ in $S$ relating the SPQ to lattice tensor
operators arise because, first,
the lattice tensor operators
${\cal O}^\text{Lat}_{T1}$ and ${\cal O}^\text{Lat}_{T2}$ 
are defined with the indices constrained to satisfy $\mu < \nu$,
rather than being freely summed as in the continuum operators,
and, second, because the continuum tensor operators come
with a factor of $4$ rather than $2$
[see Eqs.~(\ref{eq:O3SPQ}) and (\ref{eq:O4SPQ})].

The one-loop matrix elements in the SPQ theory,
with NDR$'$ regularization, are
\begin{align}
\langle {\cal O}_k^\text{SPQ} \rangle^\text{(1)}
&= Z^\text{PQB}_{k\ell} 
\langle {\cal O}_\ell^\text{SPQ} \rangle^\text{(0)} \nonumber \\
Z^\text{PQB}_{k\ell} &=
\delta_{kl} + \frac{\alpha}{4\pi}\left[
(R^{-1}\gamma^\PQ R)_{k\ell} \log(\lambda/\mu) + C_{k\ell}^\text{PQB}\right]
\,.
\label{eq:SPQ+match}
\end{align}
This is identical to Eq.~(\ref{eq:PQBmatch})
because of the exact matching between
the PQB and SPQ bases, Eq.~(\ref{eq:matchPQSPQ}).
The one-loop lattice matrix elements take the form
\begin{equation}
\langle {\cal O}_m^\text{Lat} \rangle^\text{(1)}
= 
\langle {\cal O}_m^\text{Lat} \rangle^\text{(0)}
+ \frac{\alpha}{4\pi}\left[
\widetilde\gamma_{mn} \log(a\lambda) + C_{mn}^\text{Lat}\right]
\langle {\cal O}_n^\text{Lat} \rangle^\text{(0)}
\,,
\label{eq:Latmatch}
\end{equation}
with $\widetilde\gamma$ the one-loop anomalous dimension matrix in the lattice
basis. This satisfies
$R^{-1}\gamma^\PQ R S = S\widetilde\gamma$,
which is simply the statement that the projection of $\widetilde\gamma$ onto the
basis corresponding to ${\cal O}_k^\text{SPQ}$ is regularization independent.
Equating one-loop matrix elements after transforming the lattice results
by the matrix $S$ leads to
\begin{align}
{\cal O}_k^\text{SPQ} \cong
\Big\{ S_{km} + \frac{\alpha}{4\pi} 
\Big[-(R^{-1}\gamma^\PQ R S)_{km} \log(a\mu) \nonumber \\
+ (C^\text{PQB}S)_{km} - (S C^\text{Lat})_{km} \Big]\Big\} 
{\cal O}_m^\text{Lat}
\,.
\label{eq:matchSPQLat}
\end{align}
Note that in this case ``$\cong$'' means a genuine
matching between operators. Taste factors match and
there are no restrictions on external fields.
The only provisos are that, on the right-hand side, we have dropped
lattice operators having tastes other than $\xi_5\cdot\xi_5$
and corrections proportional to powers of the lattice spacing.

\subsection{Final Matching Results}
\label{subsec:matchresult}

Combining the results (\ref{eq:matchContPQ}), (\ref{eq:matchPQ}), 
(\ref{eq:matchPQSPQ}) and (\ref{eq:matchSPQLat}) we
can now match continuum operators $Q^\Cont_2$ and
$Q^\Cont_3$ in the Dirac basis [Eqs.~(\ref{eq:Q2}) and (\ref{eq:Q3})]
to lattice operators:
\begin{align}
Q_j^\Cont &\cong 2 z_{jm} {\cal O}^\text{Lat}_m
\label{eq:matchfinal}
\\
z_{jm}&= P_{jk}\Bigg\{ (RS)_{km} + \frac{\alpha}{4\pi}
\Bigg[ - (\gamma^\PQ RS)_{km} \ln(a \mu) 
\nonumber\\
&\quad + \left(C^\text{PQA}RS - RC^\text{PQB}S\right)_{km}
\nonumber \\
&\quad + \left(RC^\text{PQB}S - RSC^\text{Lat}\right)_{km}\Bigg]\Bigg\}
\,,
\label{eq:zfinal}
\end{align}
where $j=2,3$, $k=1-4$ and $m=1-6$.
Here $P$ is a rectangular matrix projecting out the first two operators
from the four-dimensional PQA basis of Eq.~(\ref{eq:basisPQA}).
Its only non-zero elements are $P_{21}=P_{32}=1$
[corresponding to the exact matching of Eq.~(\ref{eq:matchContPQ})].
As a matrix it looks like
\begin{equation}
P = \left(\begin{array}{cccc} 
  1 & 0 & 0 & 0  \\
  0 & 1 & 0 & 0  
      \end{array}\right)
\,.
\label{eq:Pdef}
\end{equation}
We stress again that Eq.~(\ref{eq:matchfinal}) is not a true
operator matching, but rather a shorthand indicating agreement
(at one-loop order, and up to known taste factors)
between the positive parity parts of the appropriate 
kaon mixing matrix elements
defined using the external operators described above.

We also note that the contribution from the matching to operators
in the PQB basis cancels,
as can be seen from the fact that
the $RC^\text{PQB}S$ term appears with both signs. 
Cancellation is expected since this is an intermediate scheme.
We find it useful, however, to break the result up as shown.
This simplifies the calculation (as discussed in Appendix~\ref{app:sec:match}),
and is also useful conceptually.
In particular, it is the contribution from the PQA to PQB matching,
i.e. the $C^\text{PQA}RS-RC^\text{PQB}S$ term, which was not
previously accounted for in Refs.~\cite{Sharpe:1993ur,Lee:2003sk,Kim:2011pz}.
In other words, these works effectively started with the
continuum PQB basis (defining the operators using the regularization
scheme of Ref.~\cite{Gupta:1996yt}) and matched from this to lattice
operators.

It is useful to recast our final result into the notation
used in Ref.~\cite{Kim:2011pz}:
\begin{align}
z_{jm} &= b_{jm} + \frac{g^2}{(4\pi)^2}
 \Big( - \gamma_{jm} \log (\mu a) + c_{jm}\Big)\,, \\
c_{jm} &= d^\text{Cont}_{jm}
  - d^\text{Lat}_{jm} 
 - C_F I_{MF} T_{jm} \,,
\label{eq:zij}
\end{align}
with $C_F=4/3$.
Here $b_{jm}$ gives the linear relations between operators
at tree level, while $d^\Cont$ and $d^\text{Lat}$
are the finite parts of one-loop matrix
elements in continuum and lattice regularizations, respectively.
The term proportional to the matrix $T$ appears if 
one mean-field improves the lattice operators, with 
$I_{MF}$ the appropriate lattice integral.
Details are given in Ref.~\cite{Kim:2011pz}.
This contribution is, strictly speaking, 
part of $d^\text{Lat}$ but this separation
allows one to see the numerical impact of mean-field improvement.
We note that below we use the formula (\ref{eq:zij})  not only
for $j=2,3$ but also for $j=4,5$.

Comparing Eqs.~(\ref{eq:zij}) and (\ref{eq:zfinal}), we see that
\begin{eqnarray}
b_{jm} &=& \left(PRS\right)_{jm}
\\
\gamma_{jm} &=& \left(P\gamma^\PQ RS\right)_{jm}
\\
d^{\Cont}_{jm} &=& \left(PC^\text{PQA}RS\right)_{jm}
\label{eq:dCont}
\\
d^\Lat_{jm} &=& \left(PRSC^\text{Lat}\right)_{jm}-C_F I_{MF} T_{jm}
\,.
\end{eqnarray}
The anomalous dimension matrix $\gamma^\PQ$ can be obtained,
for example, from Ref.~\cite{Buras:2000if}.
The new quantity $d^{\Cont}$ is calculated in App.~\ref{app:sec:match}.
The finite part of the lattice one-loop matrix elements,
$d^\Lat$, and the mean-field improvement matrix
$T$ are calculated for our choice of operators and
action in Ref.~\cite{Kim:2011pz}.
We collect all these results in Tables~\ref{tab:matchQ2}
and ~\ref{tab:matchQ3}.

\begin{table}[tbp!] 
\caption[]{Components of the matching coefficients for $Q_2^\Cont$,
as defined in Eq.~(\ref{eq:zij}).
The $d^\text{Lat}$ are for HYP-smeared valence
fermions and operators and the Symanzik gauge action.
The last column gives the numerical values for the complete
one-loop matching coefficients $z_{2m}$ for mean-field improved
operators (for which $I_{MF}=0.722795$) 
on the MILC ultrafine lattices ($\alpha=0.2098$) 
and with $\mu a=1$.}
\label{tab:matchQ2}
\begin{center}
\begin{tabular}{lrrrrrr}
\hline
Operator $m$ & $b_{2m}$  &$\gamma_{2m}$  &$d^\text{Cont}_{2m}$  
& $d^\text{Lat}_{2m}$ & $T_{2m}$ & $z_{2m}$ \\
\hline 
$\mathcal{O}^\text{Lat}_{S1}$   
& $-1/2$ & $6$   & $-11/6$ &   2.335 & -1  & -0.554\\
$\mathcal{O}^\text{Lat}_{S2}$   
& $1$    & $-10$ & $+13/6$ & -14.528 &  6  &  1.182\\
$\mathcal{O}^\text{Lat}_{P1}$   
& $-1/2$ & $6$   & $-11/6$ &   3.174 & -1  & -0.568\\
$\mathcal{O}^\text{Lat}_{P2}$   
& $1$    & $-10$ & $+13/6$ &   4.061 & -2  &  1.001\\
$\mathcal{O}^\text{Lat}_{T1}$   
& $1/2$& $-14/3$ & $+5/6$ &   -2.518 &  1  &  0.540\\
$\mathcal{O}^\text{Lat}_{T2}$
& $0$    & $2/3$ &  $-1/2$  &  0.012 &  0  & -0.009\\
\hline
\end{tabular}
\end{center}
\end{table}

\begin{table}[tbp!] 
\caption[]{Components of the matching coefficients for $Q_3^\Cont$.
Notation as in Table~\ref{tab:matchQ2}.}
\label{tab:matchQ3}
\begin{center}
\begin{tabular}{lrrrrrr}
\hline
Operator $m$ & $b_{3m}$  &$\gamma_{3m}$  &$d^\text{Cont}_{3m}$  
& $d^\text{Lat}_{3m}$ & $T_{3m}$ & $z_{3m}$ \\
\hline 
$\mathcal{O}^\text{Lat}_{S1}$   
& $6$  & $88$  & $+50/3$ & -13.703 & 12  &  6.314\\
$\mathcal{O}^\text{Lat}_{S2}$   
& $0$  & $-40$ & $-46/3$ & -15.035 &  0  & -0.005\\
$\mathcal{O}^\text{Lat}_{P1}$   
& $6$  & $88$  & $+50/3$ & -23.769 & 12  &  6.482\\
$\mathcal{O}^\text{Lat}_{P2}$   
& $0$  & $-40$ & $-46/3$ &  15.165 &  0  & -0.509\\
$\mathcal{O}^\text{Lat}_{T1}$   
& $2$& $ 8/3$  & $-14/3$ &  -8.164 &  4  &  1.994\\
$\mathcal{O}^\text{Lat}_{T2}$
& $4$ & $136/3$ &  $+6$  & -12.376 &  8  &  4.178\\
\hline
\end{tabular}
\end{center}
\end{table}

We include in the last column of each table the numerical
values of the matching coefficients for mean-field
improved operators on the finest
MILC ensemble used in our companion numerical 
study~\cite{Bae:2013tca,Bae:2013mja}.
These are the operators we use in practice. 
Comparing the results for $z_{mj}$ to the tree-level
values, $b_{mj}$ shows that the one-loop perturbative
corrections are $\sim 5\%$.

As noted above, the matching results of Ref.~\cite{Kim:2011pz}
for the operators $Q_1^\Cont$, $Q_4^\Cont$ and $Q_5^\Cont$ 
remain valid, because the
missing PQA to PQB matching step turns out to  have
a vanishing one-loop coefficient.
This is explained in Appendix~\ref{app:subsec:otherops}.
However, since the results in Ref.~\cite{Kim:2011pz} are
presented for operators in the SUSY basis,
and also for completeness, we collect the results for
$Q_4^\Cont$ and $Q_5^\Cont$ in Tables~\ref{tab:matchQ4}
and \ref{tab:matchQ5}.
The results for the $B_K$ operator $Q_1^\Cont$ can be read
directly from Ref.~\cite{Kim:2011pz}.

\begin{table}[tbp!] 
\caption[]{Components of the matching coefficients for $Q_4^\Cont$.
Notation as in Table~\ref{tab:matchQ2}.}
\label{tab:matchQ4}
\begin{center}
\begin{tabular}{lrrrrrr}
\hline
Operator $m$ & $b_{4m}$  &$\gamma_{4m}$  &$d^\text{Cont}_{4m}$  
& $d^\text{Lat}_{4m}$ & $T_{4m}$ & $z_{4m}$ \\
\hline 
$\mathcal{O}^\text{Lat}_{S1}$   
& $0$    & $0$   &  $-3$ &       0    &  0 & -0.050\\
$\mathcal{O}^\text{Lat}_{S2}$   
& $1$    & $-16$ & $+23/3$ & -16.0196 &  6 &  1.299\\
$\mathcal{O}^\text{Lat}_{P1}$   
& $0$    & $0$   &  $+3$ &       0    &  0 &  0.050\\
$\mathcal{O}^\text{Lat}_{P2}$   
& $-1$   & $16$  & $-23/3$ &  -5.0862 &  2 & -1.075\\
$\mathcal{O}^\text{Lat}_{V1}$   
& $-1/2$ & $8$   & $-23/6$   & 2.7193 & -1 & -0.593\\
$\mathcal{O}^\text{Lat}_{V2}$
& $0$    & $0$   & $+3/2$ &    0.3401 &  0 &  0.019\\
$\mathcal{O}^\text{Lat}_{A1}$   
& $1/2$  & $-8$  & $+23/6$  & -2.9543 &  1 &  0.597\\
$\mathcal{O}^\text{Lat}_{A2}$
& $0$    & $0$   & $-3/2$ &    0.3651 &  0 & -0.031\\
\hline
\end{tabular}
\end{center}
\end{table}

\begin{table}[tbp!] 
\caption[]{
Components of the matching coefficients for $Q_5^\Cont$.
Notation as in Table~\ref{tab:matchQ2}.}
\label{tab:matchQ5}
\begin{center}
\begin{tabular}{lrrrrrr}
\hline
Operator $m$ & $b_{5m}$  &$\gamma_{5m}$  &$d^\text{Cont}_{5m}$  
& $d^\text{Lat}_{5m}$ & $T_{5m}$ & $z_{5m}$ \\
\hline 
$\mathcal{O}^\text{Lat}_{S1}$   
& $-2$  & $-4$  & $-1/3$ &   6.835 & -4 &  -2.055 \\
$\mathcal{O}^\text{Lat}_{S2}$   
& $0$   & $12$  & $1$    &   4.892 &  0 &  -0.065 \\
$\mathcal{O}^\text{Lat}_{P1}$   
& $2$   &  $4$  & $1/3$  & -10.191 &  4 &   2.111 \\
$\mathcal{O}^\text{Lat}_{P2}$   
& $0$   & $-12$ & $-1$   &   5.174 &  0 &  -0.103 \\
$\mathcal{O}^\text{Lat}_{V1}$   
& $0$   & $-6$  & $-1/2$ &  -0.537 &  0 &   0.001 \\
$\mathcal{O}^\text{Lat}_{V2}$
& $1$   & $2$   & $1/6$  &  -8.083 &  4 &   1.073 \\
$\mathcal{O}^\text{Lat}_{A1}$   
& $0$   & $6$   & $1/2$  &   0.537 &  0 &  -0.001 \\
$\mathcal{O}^\text{Lat}_{A2}$
& $-1$  & $-2$  & $-1/6$ &  -0.179 &  0 &  -1.000 \\
\hline
\end{tabular}
\end{center}
\end{table}

To complete the description of the matching results
used in Refs.~\cite{Bae:2013tca,Bae:2013mja}.
we must also consider the denominator of the
$B$-parameters defined in Eq.~(\ref{eq:def-B_i}).
The matching of the pseudoscalar bilinears in the
denominator is given by
\begin{eqnarray}
\bar s \gamma_5 d &\cong&
z_P {\cal O}_P^\Lat
\label{eq:zPmatch}
\\
{\cal O}_P^\Lat &=& 
\bar\chi_s \overline{(\gamma_5\otimes\xi_5)} \chi_d
\,,
\end{eqnarray}
where, as above, the symbol $\cong$ implies matching of matrix
elements (here connecting the vacuum to an appropriate
kaon or antikaon) up to taste factors (here a single factor
of $N_T$ since there is only one ``taste loop'').
The notation for the staggered bilinear is as in Ref.~\cite{Kim:2010fj}.
For the bilinear matrix elements we do not need to introduce the
extra $d$ and $s$ quarks of the PQQCD and SPQ theories, since there
is only one contraction. For the same reason, there is no contraction
factor of $2$ in Eq.~(\ref{eq:zPmatch})
[as compared, say, to Eq.~(\ref{eq:matchfinal})].
The one-loop result for $z_P$ is~\cite{Kim:2010fj}
\begin{equation}
z_P = 1 + \frac{\alpha}{4\pi}
\left(8 \log(\mu a) + 10/3 - 1.57938\right)
\,.
\end{equation}
This is for the continuum bilinear defined in the
NDR scheme and the lattice bilinear composed of HYP-smeared
valence fermions with Symanzik-improved glue.
Note that the lattice operator involves no gauge links and thus
cannot be mean-field improved.

In terms of lattice operators, the $B$-parameters thus become
\begin{equation}
B_i(\mu)
= \frac{2 \langle \overbar{K}^0_{P1}| z_{ij}{\cal O}_j^\Lat |K^0_{P2}\rangle}
{N_i
\langle \overbar{K}_P|z_P {\cal O}_P^\Lat|0\rangle
\langle 0|z_P {\cal O}_P^\Lat|K_0\rangle}
\,.
\end{equation}
This is now an equality (up to two-loop and discretization corrections)
since the taste factors cancel between numerator and denominator.

\section{Renormalization Group Evolution}
\label{sec:rg_evol}

In our numerical calculations, we match matrix elements of
lattice operators to those of continuum operators
using Eqs.~(\ref{eq:matchfinal}) and (\ref{eq:zfinal}).
To avoid large logarithms in this matching, we set $\mu=1/a$. 
Before taking the continuum limit,
we must evolve the resulting matrix elements to a common scale.
We do this using the most accurate anomalous dimensions available,
which in this case are of two-loop order.
Although this is a standard procedure, there are some
subtleties which arise for the operators under consideration.
In this section we discuss these subtleties.

RG evolution can be expressed as
\begin{equation}
  \langle Q_i(\mu_b)\rangle = W(\mu_b, \mu_a)_{ij} 
\langle Q_j(\mu_a)\rangle \,,
  \label{eq:Q_rg_def1}
\end{equation}
where the matrix kernel satisfies
\begin{equation}
  \frac{d}{d \ln \mu_b} W(\mu_b, \mu_a)_{ij}
  = - \gamma(\mu_b)_{ik} W(\mu_b, \mu_a)_{kj}\,,
  \label{eq:RGW}
\end{equation}
together with the boundary condition $W^Q(\mu_a,\mu_a)_{ij}=\delta_{ij}$.
We expand the anomalous dimension matrix as
\begin{align}
  \gamma(\mu) &= 
  \gamma^{(0)} \frac{\alpha(\mu)}{4\pi} 
  + \gamma^{(1)} \left(\frac{\alpha(\mu)}{4\pi}\right)^2 + \cdots\,.
\end{align}
Results for $\gamma^{(0)}$ and $\gamma^{(1)}$
for the operators of interest
are collected in Appendix~\ref{app:sec:anomdim}.
There is mixing within the operator pairs $\{Q_2^\Cont,Q_3^\Cont\}$ 
and $\{Q_4^\Cont,Q_5^\Cont\}$, while $Q_1^\Cont$ and the
pseudoscalar density do not mix. 

To evolve the $B$-parameters $B_{2-5}$,
we first take out the normalization factors $N_i$ from
Eq.~(\ref{eq:def-B_i}) by defining
\begin{eqnarray}
R_i(\mu) &\equiv& N_i B_i(\mu)
\\
&=& \frac{\langle \overbar{K}_0| Q_i^\Cont | K_0\rangle}
{\langle \overbar{K}_0|\bar s\gamma_5 d|0\rangle
\langle 0|\bar s\gamma_5 d|K_0\rangle}
\,.
\label{eq:Rdef}
\end{eqnarray}
These quantities can be run to a common scale, and then
divided by the $N_i$ to return to the $B$-parameters.
Defining $W^R(\mu_b,\mu_a)$ as the RG kernel for the $R_i$,
\begin{equation}
R_i(\mu_b) = W^R(\mu_b,\mu_a)_{ij} R_j(\mu_a)
\,,
\end{equation}
we have
\begin{equation}
W^R(\mu_b,\mu_a)_{ij} = \frac{W(\mu_b,\mu_a)_{ij}}{[W^P(\mu_b,\mu_a)]^2}
\,,
\label{eq:WRdef}
\end{equation}
where $W^P$ describes the evolution of the pseudoscalar density.
Combining these results we arrive at
\begin{align}
B_i(\mu_b) &= \sum_{j} W^B(\mu_b, \mu_a)_{ij} B_j(\mu_a) \,,
\label{eq:WB_rg_def1}
\\
W^B(\mu_b,\mu_a)_{ij} &= \frac{1}{N_i} W^R(\mu_b,\mu_a)_{ij} N_j 
\,.
\label{eq:WB_rg_def2}
\end{align}
We stress that Eqs.~(\ref{eq:Rdef}-\ref{eq:WB_rg_def2}) apply only
for the BSM operators (with $i,j=2-5$) and not for $B_K$. 
$B_K$ involves a different denominator, which does not run, so its
running, which is diagonal, is given by the element $W_{11}$
of the operator evolution kernel.

We can also consider anomalous dimensions for the $B$-parameters themselves,
defined as in Eq.~(\ref{eq:RGW}) but with $Q\to B$. This gives
\begin{equation}
  [\gamma_B]_{ij} =
  \frac{N_j}{N_i}\bigg(
  [\gamma_Q]_{ij} - 2 \gamma_P \delta_{ij}
  \bigg) \,,
\label{eq:gammaBdef}
\end{equation}
where again $i,j=2-5$.
Numerical values are given in Appendix~\ref{app:sec:anomdim}.

\subsection{Solutions for the Evolution Kernel}
\label{subsec:kernel}

The general solution of the RG equation ~\eqref{eq:RGW} is
\begin{equation}
  W(\mu_b,\mu_a) = P_\alpha
  \exp \left(
  - \int^{\alpha_b}_{\alpha_a}
  \frac{\gamma(\alpha)}{2\beta(\alpha)}d \alpha
  \right) \,,
  \label{eq:Wsoln}
\end{equation}
where $P_\alpha$ indicates ``$\alpha$-ordering'' of the matrices
in the integral, $\alpha_a=\alpha(\mu_a)$, $\alpha_b=\alpha(\mu_b)$,
and the $\beta$-function is defined with the normalization
\begin{eqnarray}
\beta(\alpha) &=& \frac12 \frac{d\alpha}{d\ln\mu}
= -\beta_0 \frac{\alpha^2}{4\pi}
-\beta_1 \frac{\alpha^3}{(4\pi)^2}
+ \cdots\,,
\end{eqnarray}
(so that $\beta_0 =9$  and $\beta_1=64$).   
%
In the literature, a standard approximate form of the general solution 
Eq.~\eqref{eq:Wsoln} is used when the anomalous dimension is known
to two-loop order~\cite{Buras:1979yt}:
\begin{align}
W&(\mu_b,\mu_a) \approx 
  \left[1\! +\! \frac{\alpha_b}{4\pi} J\right]^{-1}
  W^{(0)}(\mu_b,\mu_a)
  \left[1 \!+\! \frac{\alpha_a}{4\pi} J\right],
  \label{eq:RGWfinal1}  
\end{align}
where
\begin{align}
W^{(0)}(\mu_b,\mu_a) &= 
 V^{-1}
 \left(\frac{\alpha_b}{\alpha_a}\right)^{\gamma^{(0)}_D/2\beta_0}
 V,
\\
J &= \frac{\beta_1 \gamma^{(0)}}{2\beta_0^2}
- V^{-1} M V \,,
\label{eq:def_J}
\\
M_{ij} &= \frac{\left[V \gamma^{(1)} V^{-1}\right]_{ij}}
                {2\beta_0 + (\gamma^{(0)}_D)_{jj}-(\gamma^{(0)}_D)_{ii}}
\,.
\label{eq:def_M}
\end{align}
Here $V$ is the matrix that diagonalizes $\gamma^{(0)}$,
\begin{equation}
V \gamma^{(0)} V^{-1} = \gamma^{(0)}_D
\,,
\end{equation}
and 
%
\begin{align}
 \left[ \left(\frac{\alpha_b}{\alpha_a}\right)
  ^{\gamma^{(0)}_D / 2\beta_0} \right]_{ij}
 = \delta_{ij}\left(\frac{\alpha_b}{\alpha_a}\right)
    ^{(\gamma^{(0)}_D)_{ii} / 2\beta_0}
\,.
\end{align}
In practice, we use an alternative form of Eq.~(\ref{eq:RGWfinal1}),
\begin{align}
W(\mu_b,\mu_a) &\approx 
   W^{(0)}(\mu_b,\mu_a) 
\nonumber\\
&\ \ \!+\! \frac{1}{4\pi} \left[\alpha_a W^{(0)}(\mu_b,\mu_a) J
\!-\! \alpha_b J  W^{(0)}(\mu_b,\mu_a)\right]
 \label{eq:RGWfinal}  
\\
&\equiv
W^{(0)}(\mu_b,\mu_a) \!+\! \frac{1}{4\pi}
V^{-1} A V
\,.
 \label{eq:RGWfinal2}  
\end{align}
This form is equivalent at the order we work,
and is more convenient for the following discussion.

This approximate analytic solution fails, however, if,
for some choice of $i\ne j$,
\begin{equation}
2\beta_0+(\gamma^{(0)}_D)_{jj}-(\gamma^{(0)}_D)_{ii} = 0\,,
\label{eq:approxsolndenom}
\end{equation}
for then $M$ diverges [see Eq.~(\ref{eq:def_M})].
This indeed happens for the pair of operators $Q_{4,5}^\Cont$,
since the eigenvalues of $\gamma^{(0)}$ differ by exactly $2\beta_0=18$
(see Appendix~\ref{app:sec:anomdim}).
We stress that
this is a failure of the approximation method, and does not indicate
a breakdown in perturbative convergence for $W$ itself.
Indeed, the truncated version of the
differential equation (\ref{eq:RGW}) is not singular.

The problem can be resolved
using the analytic continuation technique introduced in 
Ref.~\cite{Adams:2007tk}.
The outcome is that,
for each $\{i,j\}$ pair for which the denominator of $M_{ij}$ vanishes
[i.e. for which Eq.~(\ref{eq:approxsolndenom}) holds],
the element $A_{ij}$ of the matrix $A$ in Eq.~(\ref{eq:RGWfinal2}) 
is replaced by
\begin{equation}
\frac{\left[V \gamma^{(1)} V^{-1}\right]_{ij}}{2\beta_0}
\alpha_b \ln\left(\frac{\alpha_b}{\alpha_a}\right)
\left(\frac{\alpha_b}{\alpha_a}\right)^{(\gamma^{(0)}_D)_{jj} / 2\beta_0}
\,.
  \label{eq:RGWfinalnew}  
\end{equation}
The derivation of this result is given in Ref.~\cite{Adams:2007tk}.

We have checked these analytic expressions by solving the
RG equation (\ref{eq:RGW}) numerically, after truncating
the anomalous dimension and $\beta$-function.
Specifically, we use the variable $t=(\ln \alpha)/(2\beta_0)$
which satisfies, at two-loop order,
\begin{equation}
\frac{dt}{d\ln\mu} = -{\frac{\alpha}{4\pi}}
\left({1+ \frac{\beta_1}{\beta_0}\frac{\alpha}{4\pi}}\right)
\,,
\end{equation}
Then
\begin{align}
\frac{d W(t_b,t_a)}{d t_b}
&=
\left(\frac{dt_b}{d\ln\mu_b}\right)^{-1}
\frac{dW(\mu_b,\mu_a)}{d \ln\mu_b}
\\
&\approx
\frac{\gamma^{(0)}\!+\! \frac{\alpha_b}{4\pi} \gamma^{(1)}}
{1\!+\! \frac{\alpha_b}{4\pi} \frac{\beta_1}{\beta_0}} W(t_b,t_a)
\\
&\approx
\left(\gamma^{(0)}\! +\! \frac{\alpha_b}{4\pi}
\left[\gamma^{(1)}\! -\! \frac{\beta_1}{\beta_0} \gamma^{(0)}\right]
\right) W(t_b,t_a)
\,,
\label{eq:RGWapprox}
\end{align}
where the approximations are allowed since they
involve dropping terms of the same
order as the missing three-loop contributions.
The resulting equation is straightforward to integrate numerically.

We find that, for the ranges over which we evolve,
the analytic and numerical results for the elements of $W$
agree to $\sim 0.01$ or better.
For example, the evolution matrix for the operators in the full Dirac
basis from $\mu_a=3\GeV$ to $\mu_b=2\GeV$ is
\begin{equation}
W_\text{anal}=
\begin{pmatrix}
1.0349 & 0 & 0 & 0 & 0 \\
0 & 0.8862 & 0.0013 & 0 & 0 \\
0 & -0.4786 & 1.1532 & 0 & 0 \\
0 & 0 & 0 & 0.8289 & 0.0106 \\
0 & 0 & 0 & 0.1310 & 1.0225 
\end{pmatrix},
\end{equation}
using the analytic results, and
\begin{equation}
W_\text{num}=
\begin{pmatrix}
1.0350 & 0 & 0 & 0 & 0 \\
0 & 0.8863 & 0.0013 & 0 & 0 \\
0 & -0.4789 & 1.1536 & 0 & 0 \\
0 & 0 & 0 & 0.8291 & 0.0105 \\
0 & 0 & 0 & 0.1308 & 1.0225 
\end{pmatrix}
\end{equation}
from the numerical solution.
Here we use $\alpha(2\GeV)=0.2959$ and $\alpha(3\GeV)=0.2448$.\footnote{%
These values are obtained at $N_f=3$ by following the four-loop running 
procedure given in Ref.~\cite{Chetyrkin:1997sg} starting from 
$\alpha(M_Z)=0.118$ with $M_Z=91187.6\MeV$.
}
The running between these two values is done using the four-loop
$\beta$-function with $N_f=3$. We use three active flavors
(despite being in the regime where the charm is active)
because this is the number of dynamical flavors in our simulations.
We use the four-loop $\beta$-function 
(despite evolving the operators using two-loop expressions)
since this incorporates some of the known higher-order terms.
Numerically this is not, however, very important.
For example, if we start from $\alpha(2\GeV)=0.2959$ and
run using the two-loop $\beta$-function we find
$\alpha(3\GeV)=0.2470$, which leads to
\begin{equation}
W_{\text{2-loop}\,\alpha}=
\begin{pmatrix}
1.0333 & 0 & 0 & 0 & 0 \\
0 & 0.8913 & 0.0012 & 0 & 0 \\
0 & -0.4563 & 1.1460 & 0 & 0 \\
0 & 0 & 0 & 0.8363 & 0.0101 \\
0 & 0 & 0 & 0.1252 & 1.0214 
\end{pmatrix}.
\end{equation}
Here we have used the numerical solution of
the evolution equation.
We see that the elements
differ by $\sim 0.02$ or less from those given above using
four-loop running of $\alpha$.

It is also interesting to see how quickly perturbation
theory is converging. This is illustrated
by comparing the matrices above to the one-loop result
\begin{equation}
W_\text{1-loop}=
\begin{pmatrix}
1.043 & 0 & 0 & 0 & 0 \\
0 & 0.900 & 0.0018 & 0 & 0 \\
0 & -0.425 & 1.126 & 0 & 0 \\
0 & 0 & 0 & 0.845 & 0 \\
0 & 0 & 0 & 0.118 & 1.021
\end{pmatrix}
\end{equation}
(obtained using the four-loop values of $\alpha$).

As a check on our calculation of $W$, we can compare
to the result for $W(3\GeV,2\GeV)$ given 
in Ref.~\cite{Mescia:2012fg}:\footnote{%
Note that Ref.~\cite{Mescia:2012fg} uses a different ordering of operators and
also quotes the transpose of $W$. Here we have converted
to our notation.
}
\begin{equation}
W_\text{MV}=
\begin{pmatrix}
1.035 & 0 & 0 & 0 & 0 \\
0 & 0.887 & 0.001 & 0 & 0 \\
0 & -0.474 & 1.152 & 0 & 0 \\
0 & 0 & 0 & 0.830 & 0.011 \\
0 & 0 & 0 & 0.130 & 1.022 
\end{pmatrix}
\end{equation}
This agrees with our results to better than
the $\pm 0.02$ variation between approximation methods,
thus checking our transcription of anomalous dimensions and
calculation of evolution matrices.

Results for the evolution kernels needed in our
numerical calculations are collected in Appendix~\ref{app:sec:num}.

\subsection{Running of ``golden'' combinations}
\label{subsec:golden}

The quantities 
\begin{equation}
B_{23}\equiv \frac{B_2}{B_3}\,,\ 
B_{45}\equiv \frac{B_4}{B_5}\,,\
B_{24}\equiv B_2 \times B_4\ 
{\rm and}\ 
B_{21}=\frac{B_2}{B_K}
\label{eq:golden}
\end{equation}
were found in Ref.~\cite{Bailey:2012wb} to have no one-loop
chiral logarithms in SU(2) chiral perturbation theory.
Thus they are expected to have better controlled chiral
extrapolations than the $B$-parameters themselves.
Following Ref.~\cite{Becirevic:2004qd},
we refer to them as ``golden'' combinations.

We are using these quantities in our companion lattice 
calculations~\cite{Bae:2013tca,Bae:2013mja}.
Indeed our central values for the $B_j$ are reconstructed
from these four golden quantities and our result for $B_K$.
Thus it is useful to have the RG running formulae directly
for the golden combinations.
The evolution of the $B$-parameters is given by
Eq.~(\ref{eq:WB_rg_def1}), with the evolution kernel $W^B$ being
non-vanishing only within the $(1)$, $(2,3)$, and $(4,5)$ blocks.
From this we can determine the evolution of the golden
combinations:
\begin{align}
B_{23}(\mu_b)
&=
\frac{W^B(\mu_b,\mu_a)_{22} B_{23}(\mu_a) +
W^B(\mu_b,\mu_a)_{23}}
{W^B(\mu_b,\mu_a)_{32} B_{23}(\mu_a) +
W^B(\mu_b,\mu_a)_{33}}\,,
\\
B_{45}(\mu_b)
&=
\frac{W^B(\mu_b,\mu_a)_{44} B_{45}(\mu_a) +
W^B(\mu_b,\mu_a)_{45}}
{W^B(\mu_b,\mu_a)_{54} B_{45}(\mu_a) +
W^B(\mu_b,\mu_a)_{55}}\,,
\\
B_{24}(\mu_b)
&=
B_{24}(\mu_a) \nonumber \\
&\times \Big({W^B(\mu_b,\mu_a)_{22} +
W^B(\mu_b,\mu_a)_{23}/B_{23}(\mu_a)}\Big)
\nonumber\\
&\times
\Big({W^B(\mu_b,\mu_a)_{44} +
W^B(\mu_b,\mu_a)_{45}/B_{45}(\mu_a)}\Big)\,,
\\
B_{21}(\mu_b) &= B_{21}(\mu_a) \nonumber \\
& \times \frac{W^B(\mu_b,\mu_a)_{22} +
W^B(\mu_b,\mu_a)/B_{23}(\mu_a)}
{W(\mu_b,\mu_a)_{11}}\,.
\label{eq:B21run}
\end{align}
Note that the running of $B_{23}$ depends only on the
initial value of this quantity, which is the case also for $B_{45}$.
For $B_{24}$, however, one needs the initial values of
$B_{24}$, $B_{23}$ and $B_{45}$,
while for $B_{21}$ one needs the initial values of both $B_{21}$
and $B_{23}$.
Note also that the denominator of Eq.~(\ref{eq:B21run}) involves
$W$ rather than $W^B$, because the denominator of $B_K$ involves
axial currents which have vanishing anomalous dimensions.

\begin{acknowledgments}
The research of W.~Lee is supported by the Creative Research
Initiatives Program (2013-003454) of the NRF grant funded by the
Korean government (MSIP).
W.~Lee would like to acknowledge the support from KISTI supercomputing
center through the strategic support program for the supercomputing
application research [No.~KSC-2013-G3-01].
The work of S.~Sharpe is supported in part by the US DOE grant
no.~DE-FG02-96ER40956.
\end{acknowledgments}


\appendix

\section{Calculation of Finite Parts of Matching Matrices}
\label{app:sec:match}

In this section we describe the calculation of the 
finite part of the continuum contribution to the one-loop matrix
elements of the operators $Q_j^\Cont$ [Eqs.~(\ref{eq:Q2}-\ref{eq:Q5})].
The main focus is on $Q_2^\Cont$ and $Q_3^\Cont$. Specifically, we determine
$C^\text{PQA}$, which is needed to determine $d^\Cont$ in Eq.~(\ref{eq:dCont}).
The calculation turns out to be simplified by
first calculating the difference $C^\text{PQA}R - RC^\text{PQB}$, 
which arises in the matching step described in Sec.~\ref{subsec:step2},
and then determining $C^\text{PQB}$.
Combining these two results we obtain $C^\text{PQA}$.

As explained in the main text, the required matching calculation
involves a change of operator basis {\em and NDR scheme}
in the PQQCD continuum theory.
Since the change of basis is, for $D=4$, 
accomplished by a Fierz transformation described by the matrix $R$,
any contribution to the one-loop matrix elements in Eqs.~(\ref{eq:PQAmatch})
and (\ref{eq:PQBmatch}) whose calculation is consistent with the
Fierz transformation will cancel in the difference
$C^\text{PQA}R - RC^\text{PQB}$.
This is, for example,
why there is no anomalous dimension term in Eq.(\ref{eq:matchPQ}).
It also means that wave function renormalization diagrams do not contribute.
The upshot is that we need only keep those parts of the one-loop diagrams
which containing $O(\epsilon)$ contributions 
{\em arising from the projections
onto the basis operators used in the two NDR schemes}
(or, equivalently, from the subtraction of evanescent operators).
These will multiply the $1/\epsilon$ pole from the loop integral,
leading to finite contributions to the matrix elements.
All other parts of the calculation are common to the two schemes and
cancel in the difference.

We call the projection-related finite contributions
$C^{\PQA,{\rm proj}}$ and $C^{\PQB,{\rm proj}}$.
We stress that they are not the complete finite
contributions, so that, e.g., $C^{\PQA,{\rm proj}}\ne C^\PQA$.
But they are the only parts we need in order to
calculate the difference
$C^\text{PQA}R - RC^\text{PQB}$.

With this background in place, we now explain, in turn,
the calculation of $C^{\PQA,{\rm proj}}$,
$C^{\PQB,{\rm proj}}$ and $C^\PQB$, from which we
obtain $C^\PQA$. We then explain why the
results for operators $Q_1^\Cont$, $Q_4^\Cont$ and $Q_5^\Cont$
from Ref.~\cite{Kim:2011pz} are not impacted by the
considerations of this appendix.

\subsection{Projection parts in PQA basis}
\label{app:subsec:PQAprojection}

For the PQA calculation, it is simplest to make a small further
change in the basis from our canonical PQA basis
(which we repeat for convenience)
\begin{equation}
\overrightarrow{{\cal O}^\text{PQA}}
= \{ Q_{2,+}^\PQ, \;Q_{3,+}^\PQ,\;Q_{2,-}^\PQ,\;Q_{3,-}^\PQ \}
\,,
\label{eq:basisPQAapp}
\end{equation}
to
\begin{equation}
\overrightarrow{{\cal O}^\text{PQ}} = 
\{Q_{2,IA}^\PQ, \;Q_{2,II}^\PQ,\;Q_{3,IA}^\PQ,\;Q_{3,II}^\PQ \}
\,,
\label{eq:basisPQ}
\end{equation}
where we recall that
\begin{equation}
Q_{j,\pm}^\PQ = Q_{j,II}^\PQ \pm Q_{j,IA}^\PQ \qquad (j=2,3)
\,.
\label{eq:QPQpmapp}
\end{equation}
We use the same definition of evanescent operators in the PQ and PQA
bases, so the two bases are exactly related by a linear transformation,
namely
\begin{equation}
{\cal O}^\PQA_k = V_{k\ell} {\cal O}^\PQ_\ell
\,,
\end{equation}
with
\begin{equation}
V = \left(\begin{array}{cccc}
1 & 1 & 0 & 0\\
0 & 0 & 1 & 1\\
-1& 1 & 0 & 0\\
0 & 0 &-1 & 1
\end{array}\right)
\,.
\end{equation}
Thus if we calculate $C^\PQ$ from
\begin{align}
\langle {\cal O}_k^\text{PQ} \rangle^\text{(1)}
& = \langle {\cal O}_k^\text{PQ} \rangle^\text{(0)}
\nonumber \\
& + \frac{\alpha}{4\pi}\left[
\gamma^\PQ_{k\ell} \log(\lambda/\mu) + C_{k\ell}^\text{PQ}\right]
\langle {\cal O}_\ell^\text{PQ} \rangle^\text{(0)}
\,,
\label{eq:PQmatch}
\end{align}
then
\begin{equation}
C^\PQA = V C^\PQ V^{-1}\,.
\end{equation}
and
\begin{equation}
C^{\PQA,{\rm proj}} = V C^{\PQ,{\rm proj}} V^{-1}\,,
\end{equation}
where $C^{\PQ,{\rm proj}}$ is the finite part of the
matrix element in the PQ basis arising from projections.

We now sketch the calculation of $C^{\PQ,{\rm proj}}$.
We illustrate the method by working in detail
through the example of matching for
${\cal O}^\PQ_1=Q_{2,IA}^\PQ = 
2 [\bar{s}_1^a L d_2^a]  [\bar{s}_2^b L d_1^b]$.
We use the terminology
that the ``Dirac structure'' of this operator is $L\cdot L$.

\begin{figure}[tbp!]
\centering
\subfigure[\ Xa ]{%
\includegraphics[width=10pc]{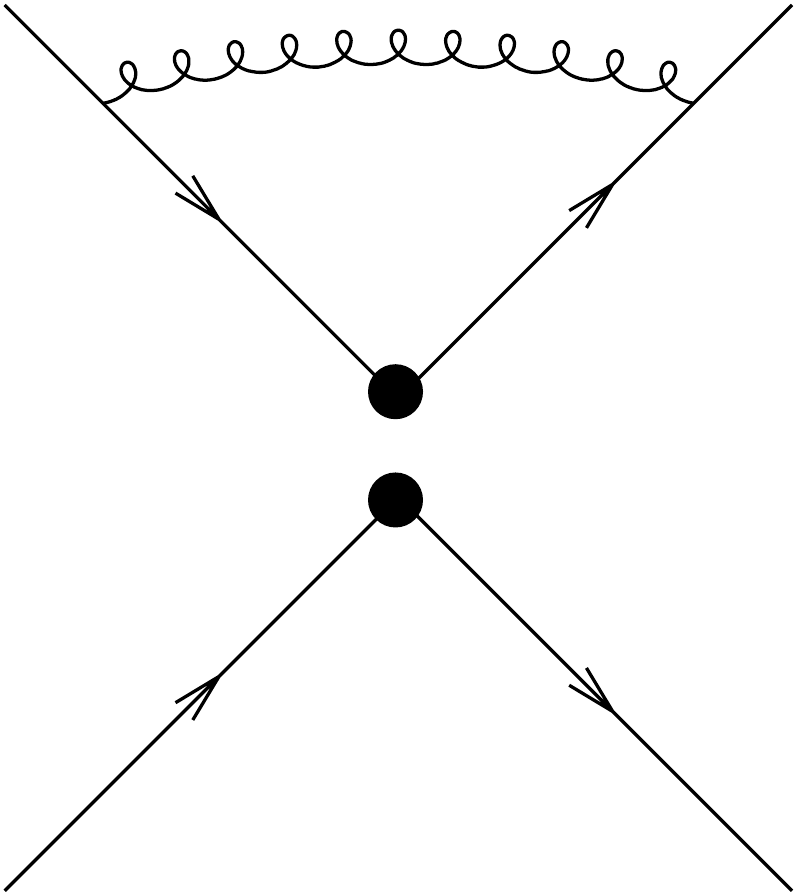}
\label{fig:Xa}}
\subfigure[\ Xb ]{%
\includegraphics[width=10pc]{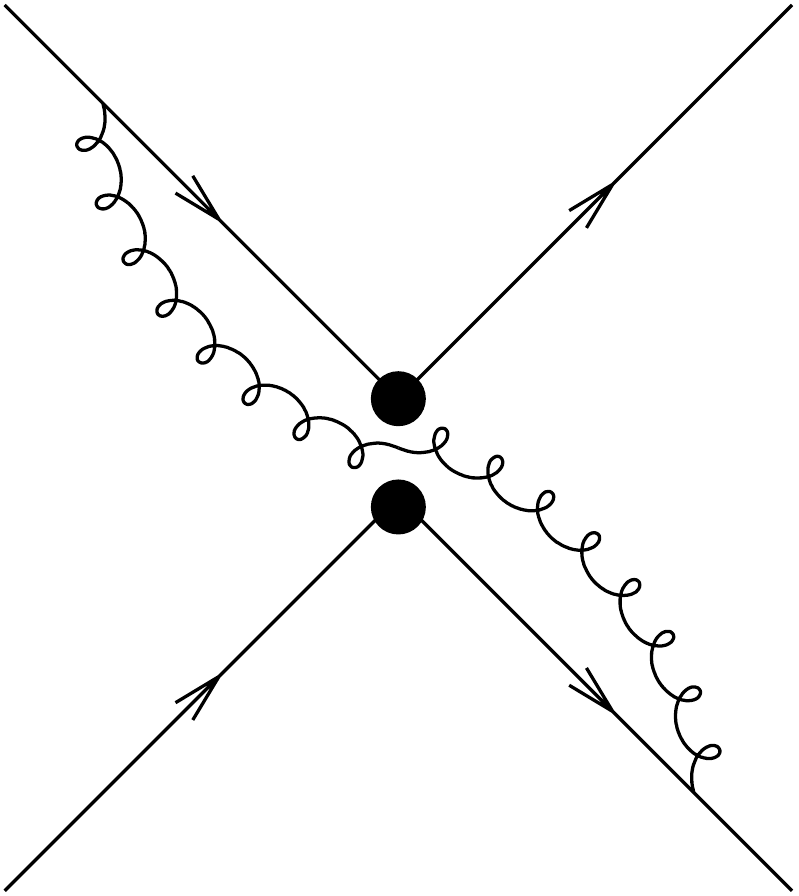}
\label{fig:Xb}}
\subfigure[\ Xc ]{%
\includegraphics[width=10pc]{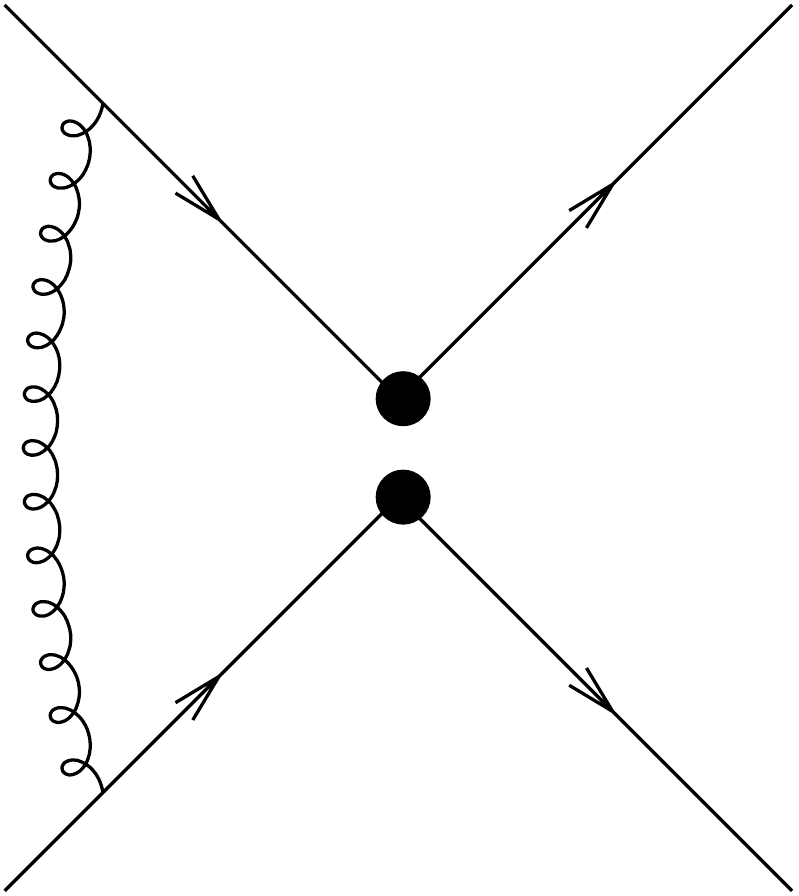}
\label{fig:Xc}}
\caption{Classes of one-loop diagrams, with labeling as
in Ref.~\cite{Patel:1992vu}. 
Each filled circle represents one of the two bilinears composing
the four-fermion operator. For each diagram shown, there is a second
one (not shown) in which the gluon connects the other two fermion
propagators.}
\label{fig:Xdiagrams}
\end{figure}

There are three types of one-loop diagrams, 
shown in Fig.~\ref{fig:Xdiagrams}.
The Xa diagrams are those in which the gluon connects
external quark and antiquark propagators which
are attached to the {\em same} bilinear.
The flavor of the bilinears depends on the operator under consideration.
For $Q_{2,IA}^\PQ$ and $Q_{3,IA}^\PQ$ [Eq.~(\ref{eq:Q3IAPQ})], 
the bilinears have flavors $\bar{s}_1 d_2$ and $\bar{s}_2 d_1$, 
while for $Q_{2,II}^\PQ$ and $Q_{3,II}^\PQ$ 
[defined in Eqs.~(\ref{eq:Q2IIPQ}) and (\ref{eq:Q3IIPQ})]
they have flavors $\bar{s}_1 d_1$ and $\bar{s}_2 d_2$. 
For $Q_{2,IA}^\PQ$, Xa diagrams have Dirac structure  
\begin{equation}
\gamma_\alpha\gamma_\beta L \gamma_\beta \gamma_\alpha \cdot L
= 16(1-\epsilon) L\cdot L\,,
\label{eq:app:XaforIA}
\end{equation}
where the $\gamma_\alpha$'s come from vertices
of the Feynman gauge gluon propagator,
while the $\gamma_\beta$'s arise from the fermion propagators
after loop integration has contracted their indices. 
The right-hand side of (\ref{eq:app:XaforIA}) is the result
after performing $D$-dimensional Dirac algebra.
In this case, the resulting operator has the same
Dirac structure as the original operator for all $D$,
so no projection is required.
The loop integral gives rise to a $1/\epsilon$ pole,
and the desired finite part is obtained from combining this
with the factor of $\epsilon$ multiplying the operator.
Taking into account the loop integral,
and the fact that there are two Xa diagrams, each giving
an identical contribution, one finds the rule that the
desired finite part is obtained by multiplying the $\epsilon$
term in the Dirac structure by $1/(2\epsilon)$
(leading to $-16\epsilon/(2\epsilon)=-8$), as well as by
the color factor. In the present case, the gluon loop simply
gives back the original color structure with an overall factor of
$C_F=4/3$. In total, then, the Xa diagrams give a 
contribution of $-32/3$ to the diagonal element $C^\PQ_{11}$. 

In the Xb diagrams the gluon connects
a quark attached to one bilinear to the antiquark attached to the other.
Thus, for $Q_{2,IA}^\PQ$, the Dirac structure is
\begin{equation}
\Gamma_b^\PQ= \gamma_\alpha\gamma_\beta L \cdot
L \gamma_\beta\gamma_\alpha
\,.
\label{eq:GammaPQb}
\end{equation}
In four dimensions one can use Fierz transformations to manipulate
this structure into a linear combination of $L\cdot L$ and
$\sigma L\cdot \sigma L$ (the latter being a shorthand for
the Dirac structure of $Q_{3}^\Cont$).
For $D\ne 4$ such manipulations introduce additional operators. 
The contribution of these evanescent operators,
which multiplies the $1/\epsilon$ pole, is then subtracted by
counterterms. What remains after this subtraction depends on the
choice of evanescent operators. This choice of scheme can be
encapsulated into rules for projecting 
Dirac structures such as $\Gamma_b^\PQ$ onto the operators
in the PQ basis.
We use the rules for the scheme of Ref.~\cite{Buras:2000if}, 
which are conveniently collected in Appendix B of Ref.~\cite{Buras:2012fs}.
In the present example, we need Eq.~(45c) from the latter work,
according to which one makes the replacement
\begin{equation}
\Gamma_b^\PQ \longrightarrow
(4-2\epsilon) L\cdot L - \sigma L \cdot \sigma L
\,.
\label{eq:diracstructure2}
\end{equation}
The desired finite part is thus
\begin{equation}
\frac{-2\epsilon L\cdot L}{2\epsilon}
= - L\cdot L\,,
\end{equation}
multiplied by the color factor.
The color factor is simple to work out but will not be needed.

Finally, we turn to the Xc diagrams, in which the gluon connects
a quark to a quark or an antiquark to an antiquark.
Here the Dirac structure is
\begin{equation}
\Gamma_c^\PQ= \gamma_\alpha\gamma_\beta L \cdot
\gamma_\alpha\gamma_\beta L
\longrightarrow (4-2\epsilon) L\cdot L+ \sigma L \cdot \sigma L
\,,
\label{eq:GammaPQc}
\end{equation}
where in the second step we have used the projection of
Eq.~(45b) of Ref.~\cite{Buras:2012fs}.
The Xc diagrams come with an additional minus sign, so
the desired finite part is obtained by multiplying the
$\epsilon$ term in Eq.~(\ref{eq:GammaPQc}) by $-1/(2\epsilon)$.
The result is 
\begin{equation}
\frac{-2\epsilon L\cdot L}{-2\epsilon}
= + L\cdot L
\,,
\end{equation}
multiplied by the {\em same} color factor as for the Xb diagrams.
Thus the contributions from the Xb and Xc diagrams cancel.

The overall result is that we know the first row of $C^{\PQ,{\rm proj}}$:
\begin{equation}
C^{\PQ,{\rm proj}}_{1k} = \left(\begin{array}{cccc} -32/3 & 0 & 0 & 0 
\end{array}\right)
\,.
\end{equation}

The calculation for the operator 
${\cal O}_2^\PQ=Q_{2,II}^\PQ=2 [\bar{s}_1^a L d_1^a] [\bar{s}_2^b L d_2^b]$
is identical. 
This is because the regularization is defined relative to the contractions of
external fields to the bilinears in the operator at hand,
irrespective of its particular flavor structure.
Since the Dirac and color structure of $Q_{2,II}^\PQ$ are the same as
that of $Q_{2,IA}^\PQ$, the results for the two operators are 
in one-to-one correspondence. The only change is that $d_1$ and $d_2$
are interchanged. The upshot is that the only non-zero entry is
$C^{\PQ,{\rm proj}}_{22}=-32/3$, so the second row of $C^{\PQ,{\rm proj}}$ is
\begin{equation}
C^{\PQ,{\rm proj}}_{2k} = \left(\begin{array}{cccc} 0 &-32/3 & 0 & 0 
\end{array}\right)
\,.
\end{equation}

We now turn to the tensor operator
${\cal O}_3^\PQ=Q_{3,IA}^\PQ=2
[\bar{s}_1^a \sigma L d_2^a][\bar{s}_2^b \sigma L d_1^b]$.
Here we are keeping the indices on $\sigma_{\mu\nu}$ implicit.
The Xa diagrams lead to 
\begin{equation}
\gamma_\alpha \gamma_\beta \sigma L \gamma_\beta\gamma_\alpha \cdot
\sigma L
\end{equation}
which vanishes through $O(\epsilon)$.
Thus there is no finite contribution from these diagrams.
For the Xb and Xc diagrams we need 
[using Eqs.~(46c) and (46b) of Ref.~\cite{Buras:2012fs}, respectively]
\begin{equation}
\gamma_\alpha\gamma_\beta \sigma L \cdot
\sigma L \gamma_\beta\gamma_\alpha
\longrightarrow
(-48+80\epsilon) L\cdot L
+(12-14\epsilon) \sigma L \cdot \sigma L
\end{equation}
and
\begin{equation}
\gamma_\alpha\gamma_\beta \sigma L \cdot
\gamma_\alpha\gamma \beta \sigma L 
\longrightarrow
(48-80\epsilon) L\cdot L
+(12-6\epsilon) \sigma L \cdot \sigma L
\,.
\end{equation}
Both types of diagram come with the same color factor, and
so can be combined. The total finite part is thus 
(remembering the relative minus sign for the Xc diagrams)
\begin{equation}
80 L\cdot L - 4 \sigma L\cdot \sigma L
\,.
\end{equation}
The diagonal color factor is $-1/6$, leading to the
results $C^{\PQ,{\rm proj}}_{31}=-80/6$ 
and $C^{\PQ,{\rm proj}}_{33}=4/6$.

There is also an off-diagonal color factor of $1/2$.
This gives rise to the combination
\begin{equation}
40 \times 2 [\bar s_1^a L d_2^b][\bar s_2^b L d_1^a]
-2 \times 2 [\bar s_1^a \sigma L d_2^b][\bar s_2^b \sigma L d_1^a]
\,.
\label{eq:tobeFierzed}
\end{equation}
Neither of these two operators is in the PQ basis (nor, for that matter,
in either of the PQA or PQB bases). To express this combination in the PQ basis
one must do a Fierz transform, which now can be done setting $D=4$ since
the matrix elements have been renormalized:\footnote{%
At ${\cal O}(\epsilon)$, this Fierz transformation introduces
further evanescent operators (which are included in the list in
Appendix A of Ref.~\cite{Buras:2000if}). These must be kept
in the calculation of two-loop anomalous dimensions.}
\begin{eqnarray}
2 [\bar s_1^a L d_2^b][\bar s_2^b L d_1^a] 
&\stackrel{D=4}{=}&
-\frac12 Q_{2,II}^\PQ + \frac18 Q_{3,II}^\PQ
\\
2 [\bar s_1^a \sigma L d_2^b][\bar s_2^b \sigma L d_1^a]
&\stackrel{D=4}{=}&
6 Q_{2,II}^\PQ + \frac12 Q_{3,II}^\PQ
\,.
\end{eqnarray}
Thus the combination in (\ref{eq:tobeFierzed}) becomes
\begin{equation}
-32 Q_{2,II}^\PQ + 4 Q_{3,II}^\PQ
\,.
\end{equation}
Combining the above we find the third row of $C^{\PQ,{\rm proj}}$ to be
\begin{equation}
C^{\PQ,{\rm proj}}_{3k} = \left(\begin{array}{cccc} -40/3 &-32 & 2/3 & 4
\end{array}\right)
\,.
\end{equation}

The calculation for the fourth row is identical to the third
aside from interchanging the roles of the two contractions,
and leads to
\begin{equation}
C^{\PQ,{\rm proj}}_{4k} = \left(\begin{array}{cccc} -32 & -40/3 & 4 & 2/3 
\end{array}\right)
\,.
\end{equation}

We can now change to the PQA basis, and find
\begin{align}
C^{\PQA,{\rm proj}}
&=V C^{\PQ,{\rm proj}} V^{-1} \nonumber \\
&= 
\frac13\left(\begin{array}{cccc}
-32 & 0 & 0 & 0\\
-136 &14 & 0 & 0\\
0& 0 & -32 & 0\\
0 & 0 & 56 & -10
\end{array}\right)
\,.
\label{eq:PQAproj}
\end{align}
In fact, we need only the first two rows,
but display the full matrix for completeness and to allow checking.

\subsection{Projection parts in PQB basis}
\label{app:subsec:PQBprojection}

We recall that we use the NDR$'$ scheme of 
Ref.~\cite{Sharpe:1993ur} in the PQB basis.
In this scheme, one uses, by definition, 
$D=4$ Fierz transforms to bring Xb and Xc diagrams into
the form of bilinear corrections. For the Xc diagrams one also
needs to charge conjugate one of the bilinears. This procedure
allows one to separate the Dirac and color parts of the
calculation. We note that this scheme is defined only at one-loop
order, but this is not a problem, both because we are working at one-loop,
and, more importantly, because we are using this scheme only as an
intermediate calculational device.

We recall that the operators in the PQB basis are
\begin{align}
{\cal O}_1^\PQB=Q_{2,I}^\PQ 
&= 2 [\bar{s}_1^a L d_1^b] [\bar{s}_2^b L d_2^a] \,,
\\
{\cal O}_2^\PQB=Q_{2,II}^\PQ 
&= 2 [\bar{s}_1^a L d_1^a] [\bar{s}_2^b L d_2^b] \,,
\\
{\cal O}_3^\PQB=Q_{3,I}^\PQ 
&= 2 [\bar{s_1}^a \sigma_{\mu\nu}L d_1^b] 
                      [\bar{s_2}^b \sigma_{\mu\nu}L d_2^a] \,,
\\
{\cal O}_4^\PQB=Q_{3,II}^\PQ 
&= 2 [\bar{s_1}^a \sigma_{\mu\nu}L d_1^a] 
                      [\bar{s_2}^b \sigma_{\mu\nu}L d_2^b] \,.
\end{align}
The Dirac structure of these four operators are the same as
those in the PQA basis. The differences between bases 
are in the flavor
indices (which has no impact since projections are
defined relative to type of contractions) and in the color
indices (which does impact the color factors).

The Xa diagrams give exactly the same finite contributions as in the
PQ basis, i.e., a factor of $-8$ for $L\cdot L$ and $0$ for 
$\sigma L\cdot\sigma L$.

For the Xb diagrams one must Fierz transform, calculate the
finite part, and then Fierz transform back. This proceeds as follows
\begin{align}
L\cdot L 
&\stackrel{\text{Fierz}}{\longrightarrow}
-\frac12 L\cdot L + \frac18 \sigma L\cdot \sigma L
\nonumber \\
&\stackrel{\text{1-loop}}{\longrightarrow}
4 L\cdot L  
\stackrel{\text{Fierz}}{\longrightarrow}
-2 L\cdot L + \frac12 \sigma L\cdot \sigma L
\end{align}
and
\begin{align}
\sigma L\cdot \sigma L 
&\stackrel{\text{Fierz}}{\longrightarrow}
6 L\cdot L + \frac12 \sigma L\cdot \sigma L
\nonumber \\
&\stackrel{\text{1-loop}}{\longrightarrow}
-48 L\cdot L 
\stackrel{\text{Fierz}}{\longrightarrow}
24 L\cdot L -6 \sigma L\cdot \sigma L
\,.
\end{align}

For the Xc diagrams there are charge conjugation steps at
the beginning and end, which
flip the sign of $\sigma L\cdot\sigma L$
while leaving $L\cdot L$ unchanged. 
Taking this into account, and including the extra sign 
from the Xc loop, one finds
\begin{eqnarray}
L\cdot L &\stackrel{\text{Xc}}{\longrightarrow}&
2 L\cdot L + \frac12 \sigma L\cdot \sigma L
\\
\sigma L\cdot\sigma L &\stackrel{\text{Xc}}{\longrightarrow}&
24 L\cdot L + 6 \sigma L\cdot \sigma L
\,.
\end{eqnarray}
Combining these results with the color factors, we find
\begin{align}
C^{\PQB,{\rm proj}} &=
\left(\begin{array}{cc} -8 & 0 \\ 0 & 0 \end{array}\right)
\otimes \left(\begin{array}{cc} -1/6 & 1/2 \\ 0 & 4/3 \end{array}\right)
\nonumber\\
&+
\left(\begin{array}{cc} -2 & 1/2 \\ 24 & -6 \end{array}\right)
\otimes
\left(\begin{array}{cc} 4/3 & 0 \\ 1/2 & -1/6 \end{array}\right)
\nonumber\\
&+
\left(\begin{array}{cc} 2 & 1/2 \\ 24 & 6 \end{array}\right)
\otimes
\left(\begin{array}{cc} -1/6 & 1/2 \\ 1/2 & -1/6 \end{array}\right)
\,,
\nonumber\\
&=\left(\begin{array}{cccc}
-5/3 & -3 & 7/12 & 1/4\\
0 & -32/3 & 1/2 & -1/6\\
28& 12 & -9 & 3\\
24 & -8 & 0 & 0
\end{array}\right)
\,.
\label{eq:CPQBproj}
\end{align}
In the tensor products the first matrix
acts on the ${Q_2,Q_3}$ indices while the second matrix
acts on the ${I,II}$ indices.

Combining this result with (\ref{eq:PQAproj}), 
we find
\begin{align}
C^\PQA -&R C^\PQB R^{-1}
\nonumber\\
&=C^{\PQA,{\rm proj}}  - R C^{\PQB,{\rm proj}} R^{-1}
\\
&= \left(\begin{array}{cccc}
-3 & -1/12 & 0 & 0 \\
-76/3 & 5/3 & 0 & 0 \\
0 & 0 & 3 & 5/12 \\
0 & 0 & 44/3 & -1/3 
\end{array}\right)
\,.
\label{eq:PQA-PQB}
\end{align}

\subsection{Finite part in PQB basis.}
\label{app:subsec:PQBfinite}

The final ingredient we need is the full finite part for
the PQB-basis operators in the NDR$'$ scheme.
The calculation proceeds essentially as in the previous
subsection, except that now we use the full finite
parts for bilinears in the NDR scheme, which can
be taken, e.g., from Ref.~\cite{Patel:1992vu}.
The method is explained in more detail in Ref.~\cite{Gupta:1996yt}.
The result is 
\begin{align}
&C^\PQB =
\left(\begin{array}{cc} 2 c_S & 0 \\ 0 & 2 c_T \end{array}\right)
\otimes \left(\begin{array}{cc} -1/6 & 1/2 \\ 0 & 4/3 \end{array}\right)
\nonumber\\
&+
\left(\begin{array}{cc} (c_S\!+\!3c_T)/2 & (c_T\!-\!c_S)/8 \\ 
                        6(c_T\!-\!c_S) &  (3c_S\!+\!c_T)/2 \end{array}\right)
\otimes
\left(\begin{array}{cc} 4/3 & 0 \\ 1/2 & -1/6 \end{array}\right)
\nonumber\\
+& 
\left(\begin{array}{cc} -(c_S\!+\!3c_T)/2 & (c_T\!-\!c_S)/8 \\ 
                        6(c_T\!-\!c_S) &  -(3c_S\!+\!c_T)/2 \end{array}\right)
\otimes
\left(\begin{array}{cc} -1/6 & 1/2 \\ 1/2 & -1/6 \end{array}\right)
\,,
\end{align}
with $c_S=2.5$ and $c_T=0.5$.
Numerically, the result takes its simplest form after a
similarity transform with $R$:
\begin{equation}
R C^\PQB R^{-1}
=
\left(\begin{array}{cccc}
31/6 & -1/24 & 0 & 0 \\
10 & -1/6 & 0 & 0 \\
0 & 0 & 49/6 & 5/24 \\
0 & 0 & -2 & 17/6 
\end{array}\right)\,.
\label{eq:app:CPQB}
\end{equation}

\subsection{Final result for $C^\PQA$}
\label{app:subsec:PQAfinal}

Combining Eqs.~(\ref{eq:PQA-PQB}) and (\ref{eq:app:CPQB}) 
we find
\begin{equation}
C^\PQA 
=
\left(\begin{array}{cccc}
13/6 & -1/8 & 0 & 0 \\
-46/3 & 3/2 & 0 & 0 \\
0 & 0 & 67/6 & 5/8 \\
0 & 0 & 38/3 & 5/2 
\end{array}\right)\,.
\label{eq:app:CPQA}
\end{equation}
Multiplying from the right by $RS$ leads to the
results quoted in Tables~\ref{tab:matchQ2} and \ref{tab:matchQ3}.

\subsection{Other Operators}
\label{app:subsec:otherops}

We have claimed above that the results given in Ref.~\cite{Kim:2011pz} for
the matching of operators $Q_{1,4,5}^\Cont$
are correct, although the matching was not done completely correctly.
Here we substantiate this claim.

We begin by discussing the $B_K$ operator,
$Q_1^\Cont$ [see Eq.~(\ref{eq:Q1})], for which the
analysis is simplest.
First we match this operator into PQQCD,
as in Sec.~\ref{subsec:step1} in the main text.
There is an exact matching of matrix elements with those of
\begin{eqnarray}
Q_1^\PQA &=& Q_{1,II}^\PQ + Q_{1,IA}^\PQ
\\
Q_{1,II}^\PQ &=& 2 [\bar s_1^a\gamma_\mu L d_1^a] 
                   [\bar s_2^b\gamma_\mu L d_2^b]
\\
Q_{1,IA}^\PQ &=& 2 [\bar s_1^a\gamma_\mu L d_2^a] 
                   [\bar s_2^b\gamma_\mu L d_1^b]
\,.
\end{eqnarray}
This forms the one-dimensional PQA basis in this case.
In Refs.~\cite{Sharpe:1993ur,Lee:2003sk,Kim:2011pz} 
it was implicitly assumed that matrix elements of
this operator are equal at one-loop order to those of
the following operator in the PQB basis
\begin{equation}
Q_1^\PQB = Q_{1,II}^\PQ 
           + 2 [\bar s_1^a\gamma_\mu L d_1^b] 
               [\bar s_2^b\gamma_\mu L d_2^a]
\,,
\end{equation}
as long as one uses the {\em same} NDR scheme for both operators.
In other words, it was assumed that $D=4$ Fierz transforms in
PQQCD commute with the calculation of one-loop corrections.
This is not valid in general. 
However, it is correct in this case, 
when using the projectors of Ref.~\cite{Buras:2000if}.
This we have checked by explicit calculation,
using the method of Sec.~\ref{app:subsec:PQAprojection}.\footnote{%
We stress that this PQA-PQB matching is different from that
just discussed for $Q_2^\Cont$ and $Q_3^\Cont$.
Here we are using the scheme of Ref.~\cite{Buras:2000if} for both bases,
while in Secs.~\ref{app:subsec:PQAprojection}
and \ref{app:subsec:PQBprojection} we use the scheme of
Ref.~\cite{Buras:2000if} in the PQA basis and the NDR$'$ scheme in
the PQB basis.}

Given this result, the PQA-PQB matching can be replaced by 
matching $Q_1^\PQB$ regularized in the scheme of Ref.~\cite{Buras:2000if}
to the {\em same} operator in the NDR$'$ scheme.
It was this latter calculation that was done (correctly) in
Refs.~\cite{Sharpe:1993ur,Lee:2003sk,Kim:2011pz}.

The same result holds true for the operators $Q_{4,5}^\Cont$:
Fierzing in the PQ theory commutes with calculating the finite
correction at one-loop (as long as one uses the same NDR scheme).
Specifically, these operators are exactly matched in PQQCD to
\begin{align}
Q_4^\PQA &=
 2 \Big\{ 
   [\bar{s}_1^a L d_1^a] 
   [\bar{s}_2^b R d_2^b] 
+  [\bar{s}_1^a L d_2^a] 
   [\bar{s}_2^b R d_1^b] 
   \Big\}\,,
\\
Q_5^\PQA &=
 2 \Big\{ 
   [\bar{s}_1^a \gamma_\mu L d_1^a] 
   [\bar{s}_2^b \gamma_\mu R d_2^b] 
+  [\bar{s}_1^a \gamma_\mu L d_2^a] 
   [\bar{s}_2^b \gamma_\mu R d_1^b] 
   \Big\}\,,
\end{align}
The claim is that, at one loop, these operators are matched
with no finite corrections to
\begin{align}
Q_4^\PQB &=
 2 \Big\{ 
   [\bar{s}_1^a L d_1^a] 
   [\bar{s}_2^b R d_2^b] 
-\frac12
   [\bar{s}_1^a \gamma_\mu L d_1^b] 
   [\bar{s}_2^b \gamma_\mu R d_2^a] 
   \Big\}\,,
\\
Q_5^\PQB &=
 2 \Big\{ 
   [\bar{s}_1^a \gamma_\mu L d_1^a] 
   [\bar{s}_2^b \gamma_\mu R d_2^b] 
-2 [\bar{s}_1^a L d_1^b] 
   [\bar{s}_2^b R d_2^a] 
   \Big\}
\,,
\end{align}
as long as the regularization of 
Ref.~\cite{Buras:2000if} is used in both cases.
This was implicitly assumed in
Refs.~\cite{Sharpe:1993ur,Lee:2003sk,Kim:2011pz}.
Because this assumption is correct,
the matching calculations done in these works 
remain valid.
We have double-checked this by repeating the calculation
from scratch.

This result does {\em not hold}, however, for $Q_{2,3}^\Cont$.
Fierzing does not commute with one-loop correcting
when using the scheme of Ref.~\cite{Buras:2000if}
in both PQA and PQB bases.\footnote{%
To see this requires an additional calculation from that presented
above, since the difference quoted above is due both to the basis
change and the change in NDR scheme.} 

\section{Anomalous dimensions}
\label{app:sec:anomdim}

We collect here the anomalous dimensions needed to evolve
the $B$-parameters of Eq.~(\ref{eq:def-B_i}) and
the golden ratios discussed in Sec.~\ref{subsec:golden}.
All anomalous dimensions are in the NDR scheme,
with those for the four-fermion operators using the
choices of evanescent operators given in Ref.~\cite{Buras:2000if}.

The two-loop anomalous dimension matrices
for $Q^\text{Cont}_{2,3}$ operators are 
calculated in Ref.~\cite{Buras:2000if}. (They are the 
same as for the $\mathcal{Q}^\text{SLL}_{1,2}$ of that work,
since the operators differ only by an overall factor.)
For $N_c=3$ and $N_f=3$, the results are
\begin{eqnarray}
\gamma_{LL}^{(0)} &=& \left(\begin{array}{cc}
-10 & 1/6 \\ -40 & 34/3 
\end{array}\right) \,,
\\
\gamma_{LL}^{(1)} &=& \left(\begin{array}{cc}
-1237/9 & -37/36 \\ -4580/9 & 557/3 
\end{array}\right) \,.
\end{eqnarray}
The eigenvalues for $\gamma_{LL}^{(0)}$ are $11.0161$ and $-9.68278$.

For $Q^\text{Cont}_{4,5}$, the anomalous dimensions are the same as for
$\mathcal{Q}^\text{LR}_{2,1}$ of Ref.~\cite{Buras:2000if}. 
Taking into account that our ordering of the operators is opposite to 
that in Ref.~\cite{Buras:2000if}, we have
\begin{eqnarray}
\gamma_{LR}^{(0)} &=& \left(\begin{array}{cc}
-16 & 0 \\ 12 & 2 
\end{array}\right) \,,
\\
\gamma_{LR}^{(1)} &=& \left(\begin{array}{cc}
-1207/6 & 201/4 \\ 154 & 49/3 
\end{array}\right) \,.
\end{eqnarray}
The eigenvalues of $\gamma_{LR}^{(0)}$ are $-16$ and $2$.

The anomalous dimension of the pseudoscalar density 
(which is the opposite of that of the quark mass) has 
coefficients~\cite{Tarrach:1980up}
\begin{equation}
  \gamma_P^{(0)} = -8\,, \qquad
  \gamma_P^{(1)} = -\frac{364}{3}\,. 
\end{equation}
For the golden combinations, we also need the anomalous dimension
of the $B_K$ operator $Q_1^\Cont$, which 
has coefficients~\cite{Buras:1989xd}
\begin{equation}
  \gamma^{(0)} = 4\,, \qquad
  \gamma^{(1)} = -17/3\,. 
\label{eq:anom_dim_bk}
\end{equation}

Finally, we can use these results in Eq.~(\ref{eq:gammaBdef})
to obtain the two-loop anomalous dimensions of the $B$-parameters
themselves. For $B_{2,3}$ we find
\begin{eqnarray}
\gamma^{(0)}_{BLL} &=& \left(\begin{array}{cc}
6 & 2/5 \\ -50/3 & 82/3 
\end{array}\right) \,,
\label{eq:gB0_SLL}\\
\gamma^{(1)}_{BLL} &=& \left(\begin{array}{cc}
947/9 & -37/15 \\ -5725/27 & 1285/3
\end{array}\right)\,, 
\label{eq:gB1_SLL}
\end{eqnarray}
while the results for $B_{4,5}$ are
\begin{eqnarray}
\gamma^{(0)}_{BLR} &=& \left(\begin{array}{cc}
0 & 0 \\ -18 & 18 
\end{array}\right) \,,
\label{eq:gB0_LR} \\
\gamma^{(1)}_{BLR} &=& \left(\begin{array}{cc}
83/2 & -67/2 \\ -231 & 259 
\end{array}\right) \,.
\label{eq:gB1_LR}
\end{eqnarray}

\section{Numerical Results for Evolution Kernels}
\label{app:sec:num}

In our numerical simulations we require
the evolution kernels to run the $B$-parameters we evaluate at
the lattice scales, $1/a$, to a canonical scale.
We use MILC collaboration asqtad ensembles~\cite{Bazavov:2009bb}
having four nominal lattice spacings.
These are labeled
$C$, $F$, $S$ and $U$ for coarse, fine, superfine and ultrafine, 
respectively.
Strictly speaking, the lattice spacings vary slightly within
the coarse ensembles, and similarly for the fine and superfine ensembles.
Here we choose a representative ensemble at each nominal lattice spacing. 
These are, in the notation of Ref.~\cite{Bae:2013tca}, 
the C3, F1, S1 and U1 ensembles,
all of which have sea quarks in the ratio $m_\ell/m_s = 1/5$.
In our numerical work, we evaluate the kernels
using the appropriate lattice spacing for each ensemble.

The inverse lattice spacings and corresponding coupling constants are
\begin{align}
 a_{C}^{-1} &= 1.657 \GeV\,,\qquad \alpha(a_{C}^{-1}) = 0.3291
 \label{eq:alpha_coarse}
 \\ 
 a_{F}^{-1} &= 2.342 \GeV\,,\qquad \alpha(a_{F}^{-1}) = 0.2734
 \label{eq:alpha_fine}
 \\ 
 a_{S}^{-1} &= 3.353 \GeV\,,\qquad \alpha(a_{S}^{-1}) = 0.2340
 \label{eq:alpha_superfile}
 \\ 
 a_{U}^{-1} &= 4.504 \GeV\,,\qquad \alpha(a_{U}^{-1}) = 0.2098
\label{eq:alpha_ultrafine}
\end{align}
These lattice spacings are obtained
from the results for the mass-dependent $r_1/a$ and using
$r_1=0.3117\;$fm~\cite{Bazavov:2009bb,Bernard}.
The coupling constants are in the $\overline{\text MS}$ scheme,
and are obtained using four-loop running as described
in Sec.~\ref{subsec:kernel}.

%

We take the canonical final scale to be either $2\GeV$,
the traditional value, 
or $3\GeV$, which is used, for example, in Ref.~\cite{Boyle:2012qb}.
The values of $\alpha$ at these scales are given
in Sec.~\ref{subsec:kernel}.
We calculate the
evolution kernel assuming $N_f=3$, although
some of our scales are higher than the charm mass.
This is appropriate because our simulations have $N_f=2+1$ flavors
of dynamical quarks.

Results for the evolution kernel for the
$B$-parameters, i.e. $W^B(\mu_f,\mu_i)$ of Eq.~(\ref{eq:WB_rg_def2}),
are given in Tables~\ref{tab:BKK}, \ref{tab:BLL} and \ref{tab:BLR}.
These are obtained using numerical integration of the two-loop RG equations,
using the method described in Sec.~\ref{subsec:kernel}.
The elements of these kernels
agree within $\sim 0.01$ with those obtained using
the analytic expressions described in Sec.~\ref{subsec:kernel},
and to within $\sim 0.02$ with those obtained
using two-loop running for $\alpha$.

\begin{table}[htbp!] 
\caption[]{
Results for evolution kernel for $B_K$, $W_{11}(\mu_f,\mu_i)$. 
Note that this is the same as the kernel for
the operator $Q_1^\Cont$.}
\label{tab:BKK}
\begin{center}
\begin{tabular}{l c c}
\hline
\vphantom{\bigg|} 
$\mu_i$ & $W_{11}(2\GeV,\mu_i)$ & $W_{11}(3\GeV,\mu_i)$ \\
\hline
\\[-0.2cm]
$a_C^{-1}$ & 0.982 & 0.948
\\
$a_F^{-1}$ & 1.014 & 0.980 
\\
$a_S^{-1}$ & 1.044 & 1.008
\\
$a_U^{-1}$ & 1.065 & 1.030
\\
\hline
\end{tabular}
\end{center}
\end{table}

\begin{table}[tbp!] 
\caption[]{
Evolution matrices, $W^B_{LL}(\mu_f,\mu_i)$,
for $B$-parameters of LL operators $\{B_2,B_3\}$.}
\label{tab:BLL}
\begin{center}
\begin{tabular}{lp{3cm}p{3cm}}
\hline
\vphantom{\bigg|} 
$\mu_i$\ \ & $W^B_{LL}(2\GeV,\mu_i)$ & $W^B_{LL}(3\GeV,\mu_i)$ \\
\hline
\\[-0.2cm]
$a_C^{-1}$\ \ &  
$\begin{pmatrix} 0.956 & -0.001 \\  0.100 & 0.822 \end{pmatrix}$ & 
$\begin{pmatrix} 0.885 & -0.003 \\  0.224 & 0.584 \end{pmatrix}$ 
\\[0.3cm]
$a_F^{-1}$\ \ &  
$\begin{pmatrix} 1.033 & 0.001 \\  -0.090 & 1.154   \end{pmatrix}$ & 
$\begin{pmatrix} 0.956 & -0.002 \\  0.101 & 0.821   \end{pmatrix}$ 
\\[0.3cm]
$a_S^{-1}$\ \ &  
$\begin{pmatrix} 1.100 & 0.005 \\ -0.316 & 1.522 \end{pmatrix}$ & 
$\begin{pmatrix} 1.018 & 0.001 \\ -0.048 & 1.083 \end{pmatrix}$ 
\\[0.3cm]
$a_U^{-1}$\ \ &  
$\begin{pmatrix} 1.147 & 0.008 \\ -0.519 & 1.840 \end{pmatrix}$ & 
$\begin{pmatrix} 1.063 & 0.003 \\ -0.186 & 1.310 \end{pmatrix}$
\\[0.3cm]
\hline
\end{tabular}
\end{center}
\end{table}

\begin{table}[htbp!] 
\caption[]{
Evolution matrices, $W^B_{LR}(\mu_f,\mu_i)$,
for $B$-parameters of LR operators $\{B_4,B_5\}$.}
\label{tab:BLR}
\begin{center}
\begin{tabular}{l p{3cm}p{3cm}}
\hline
\vphantom{\bigg|} 
$\mu_i$ \ \ \ & $W^B_{LR}(2\GeV,\mu_i)$ & $W^B_{LR}(3\GeV,\mu_i)$ \\
\hline
\\[-0.2cm]
$a_C^{-1}$ \ \ \ &  
$\begin{pmatrix} 0.994 & 0.005 \\   0.114 & 0.882 \end{pmatrix}$ &
$\begin{pmatrix} 0.986 & 0.011 \\   0.281 & 0.710 \end{pmatrix}$ 
\\[0.3cm]
$a_F^{-1}$ \ \ \ &  
$\begin{pmatrix} 1.004 & -0.004 \\ -0.094 & 1.097 \end{pmatrix}$ &
$\begin{pmatrix} 0.995 & 0.004 \\   0.116 & 0.881 \end{pmatrix}$ 
\\[0.3cm]
$a_S^{-1}$ \ \ \ &  
$\begin{pmatrix} 1.013 & -0.011 \\ -0.304 & 1.312 \end{pmatrix}$ &
$\begin{pmatrix} 1.002 & -0.002 \\ -0.051 & 1.053 \end{pmatrix}$ 
\\[0.3cm]
$a_U^{-1}$ \ \ \ &  
$\begin{pmatrix} 1.019 & -0.016 \\ -0.473 & 1.485 \end{pmatrix}$ &
$\begin{pmatrix} 1.007 & -0.006 \\ -0.186 & 1.191 \end{pmatrix}$
\\[0.3cm]
\hline
\end{tabular}
\end{center}
\end{table}

\bibliography{ref} 

\end{document}